\documentclass[3p]{elsarticle}
\usepackage{lineno,hyperref}
\usepackage{subcaption}
\usepackage{amsmath}
\usepackage{forest}
\usetikzlibrary{shapes,arrows,positioning,automata}

\tikzset{
  treenode/.style = {shape=rectangle, rounded corners,
                     draw, align=center,
                     top color=white,
                     bottom color=white},
  root/.style     = {treenode, font=\Large,
                     bottom color=red!30},
  env/.style      = {treenode, font=\ttfamily\normalsize},
  dummy/.style    = {circle,draw}
}
\usepackage{caption}
\usepackage{threeparttable}
\usepackage{algorithm,algpseudocode}
\usepackage[usestackEOL]{stackengine}
\usepackage{xcolor}
\edef\tmp{\the\baselineskip}
\usepackage{amssymb}
\newcommand*{\QEDA}{\ensuremath{\square}}
\setstackgap{L}{\tmp}
\modulolinenumbers[1]

\journal{Journal}

\begin{document}







\begin{frontmatter}

\title{Seismological Understanding of Accelerogram Amplitude Scaling for Engineers with Implications to Seismic Risk Analysis}

\author{Somayajulu L. N. Dhulipala\corref{mycorrespondingauthor}}
\address{}\cortext[mycorrespondingauthor]{Corresponding author; Email: lakshd5@vt.edu}


\begin{abstract}
Due to the paucity of strong recorded accelerograms, earthquake engineering analysis relies on accelerogram amplitude scaling for structural damage/collapse assessment and target spectrum matching. This paper investigates seismological characteristics of scaled accelerograms so as to inform future ground motion selection and seismic risk assessment methods. If a recorded accelerogram is scaled linearly by multiplying it with a positive factor, it is shown using the Representation theorem and the accelerogram Fourier spectrum that moment magnitude scales logarithmically and static stress drop scales linearly. Other seismic parameters such as the Joyner-Boore distance metric and the effective rupture area are invariant to scaling an accelerogram. This proposed interpretation of scaling is validated in the time as well as the frequency domains using a hybrid method for ground motion simulation. Finally, a discussion is made over the seismological correctness of accelerogram scaling. It is suggested that a suite of scaled accelerograms can be considered as being seismologically correct if this suite's magnitude given rupture area and stress drop distributions are similar to empirical observations.
\end{abstract}


\end{frontmatter}


\section{Introduction}\label{Intro}

\begin{figure}[h]
\begin{centering}
\includegraphics[width=0.5\textwidth]{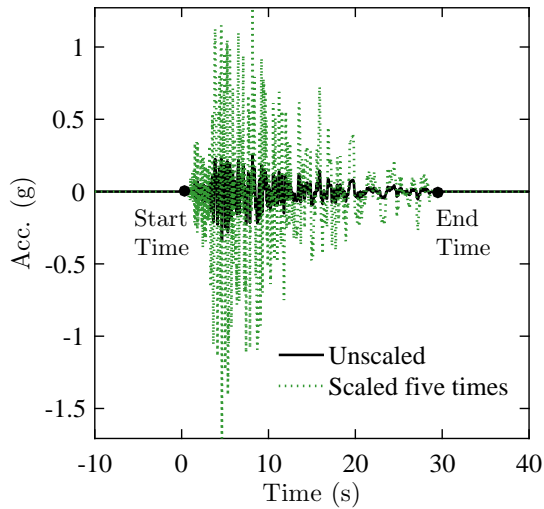}
\caption{Comparison of unscaled and scaled accelerograms from the Northridge earthquake recorded at Saticoy station. While the scaled accelerogram has amplitudes five times as the unscaled one, both accelerograms are dependent on time in the same manner and have a zero phase shift between them.}\label{fig:scaled_time}
\end{centering}
\end{figure}

Accelerogram amplitude scaling entails multiplying a recorded accelerogram by a scale factor $\lambda \in (0, +\infty)$ so as to intensify the amplitudes without changing neither the duration nor the frequency content of the recording. Figure \ref{fig:scaled_time} demonstrates this scaling procedure from which it is observed that both recorded (i.e., unscaled) and scaled accelerograms have the same start and end times with zero phase shift between their amplitudes. Why are recorded accelerograms scaled? Researchers and engineers intend to assess infrastructure performance under extreme accelerograms that have the potential to inflict structural damage and to cause structural collapse; however, such extreme recordings are scarce \cite{Dhulipala2019_c, Dhulipala2018_c, Dhulipala2020_e}. Therefore, accelerogram scaling is practiced to assess the damage and collapse capacities of structures \citep{Vamvatsikos2002a,Jalayer2003, Dhulipala2020_d,Dhulipala2018_b,Dhulipala2018_a,Dhulipala2018_f} and to select appropriate ground motions for seismic structural response analysis that match with a target response spectrum \citep{Baker2006,ASCE16}. Such ground motion selection is also extremely important for assessing the seismic resilience of structures \cite{Dhulipala2021_c,Dhulipala2021_b,Dhulipala2021_a,Dhulipala2020_a,Flint2016_c}, and particularly, critical infrastructures like nuclear power plants \cite{Dhulipala2022_a, Veeraraghavan2021_a, Dhulipala2021_f}.

Research on accelerogram amplitude scaling can be divided into two camps. One camp focuses on procedures to more efficiently scale and select accelerograms that match a target response spectrum (for e.g., see \cite{Jayaram2011}). The other camp focuses on the assessment of bias in nonlinear structural seismic responses by comparing results from unscaled and scaled accelerograms (for e.g., see \cite{Luco2007f}). In contrast, this paper takes a step back and focuses on the seismological characteristics of scaled accelerograms. The intent for conducting such an investigation is to map the scale factor $\lambda$ to seismic variables such as magnitude, distance, rupture area, stress drop, and corner frequency\footnote{Loosely speaking, static stress drop may be regarded as the amount of stress released after an earthquake. Corner frequency separates the very low and the medium-high frequency parts of an accelerogram Fourier spectrum.}. This would then, in the future, enable researchers to compare these seismic variables of the scaled accelerograms with empirical observations to ascertain the seismological bias in structural responses, and further, to develop algorithms or procedures for scaling accelerograms in a seismologically consistent manner. 

Scaled accelerograms represent potential earthquake events that are yet to be realized. The ground motions resulting from such earthquake events (or, in general, any earthquake event) are governed by the Representation theorem in Seismology \citep{Aki2002}. This theorem combines the source and the path effects and mathematically describes the ground motion at a site given rupture over a fault plane. Complementing the Representation theorem is the Fourier spectrum of an accelerogram, the models for which aim to describe the frequency content and the duration characteristics of this accelerogram using parameters such as stress drop and corner frequency \citep{Brune1970}. Because the Representation theorem and the Fourier spectrum provide a complete picture of an accelerogram, these are used as means to understand the seismology of accelerogram scaling.

In the next section, the Representation theorem is used to derive the magnitude, distance, and rupture area values for a scaled accelerogram. Following this, accelerogram Fourier spectrum is used for linking the scale factor $\lambda$ to seismic parameters such as corner frequency and static stress drop. Ground motion simulations are used to then validate the proposed interpretation of scaling. Finally, a discussion is made on seismological correctness of accelerogram scaling along with its applicability to multiple sites. It should be noted that accelerogram scaling also scales the amplitudes of velocity and displacement records, Fourier spectrum, and response spectrum linearly. This is apparent conclusion will be implicitly used in the sections that follow. Additionally, scaling in the context of this paper implies accelerogram amplitude scaling as presented in Figure \ref{fig:scaled_time}.

\section{Insights from the Representation Theorem}

The Representation theorem connects rupture on the fault plane to displacement at a site. It is the cornerstone of seismology providing theoretical as well as observational insights into earthquake rupture processes and their effects. Computational techniques to synthesize accelerograms such as the Empirical Green's function method \citep{Irikura1986} or the UCSB technique \citep{Crempien2015} are developed on the basis of this theorem. In this subsection, a theoretical investigation into accelerogram scaling is carried out by revisiting the different terms in the Representation theorem and then understanding how these terms behave upon scaling a recorded ground motion.

\begin{figure}[h]
\centering
\includegraphics[width=0.52\textwidth,height=2.6in]{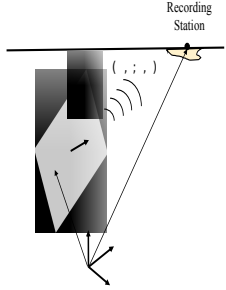}
\caption{Schematic of the earthquake process depicting rupture on the fault-plane, propagation of seismic waves, and reception at a recording station.}\label{fig:schem-pwavedens_1}
\end{figure}

Let $\mathbf{X}$ and $\pmb{\xi}$ be position vectors describing locations on the ground and the fault plane, respectively, with respect to a common coordinate system (also refer to Figure \ref{fig:schem-pwavedens_1}). Let $t$ and $\tau$ be variables keeping track of time on the ground and the fault plane, respectively. The Representation theorem for a scaled ground motion in the absence of body forces and traction is \citep{Aki2002}:

\begin{equation}
U_i^\lambda(\mathbf{X},t) = \lambda~U_i(\mathbf{X},t) = \lambda \int_{\tau}d\tau \iint_{\Sigma} \big[U_j(\pmb{\xi},\tau)\big]~c_{jkpq}~\mathbf{G}_{ip,q}(\mathbf{X},t;\pmb{\xi},\tau)~\nu_{k}~d\Sigma(\pmb{\xi})
\label{eqn:RepTheorem}
\end{equation}

\noindent
where $U_i^\lambda(\mathbf{X},t)$ is the scaled ground displacement in direction $i$\footnote{Scaling an accelerogram by a factor $\lambda$ also scales the ground displacement by $\lambda$.}, $\big[U_j(\pmb{\xi},\tau)\big]$ is the displacement discontinuity across the fault plane $\Sigma$, $c_{jkpq}$ are the elastic moduli which can depend on $\pmb{\xi}$, $\mathbf{G}_{ip,q}(\mathbf{X},t;\pmb{\xi},\tau)$ is the Green's function differentiated by $\xi_q$, $\nu_k$ is the unit normal to $\Sigma$. It is noted that $U_i(\mathbf{X},t)$ and the terms in its expansion $(\big[U_j(\pmb{\xi},\tau)\big], c_{jkpq}, \mathbf{G}_{ip,q}(\mathbf{X},t;\pmb{\xi},\tau), \textnormal{and}~\nu_{k})$ correspond to ground displacement of an unscaled accelerogram resulting from rupture on a fault plane $\Sigma$. The displacement discontinuity $\big(\big[U_j(\pmb{\xi},\tau)\big]\big)$ describes the rate of relative slip between the two faces of a fault plane as a function of space and time. As only a portion of the fault plane might experience rupture during an earthquake, it is expected that $\big[U_j(\pmb{\xi},\tau)\big]$ is not non-zero on the entire fault plane and the area within which $\big[U_j(\pmb{\xi},\tau)\big]$ is non-zero defines the rupture area. $c_{jkpq}$ represents a collection of $81$ elastic coefficients that can be a function of spatial location on the fault plane under non-homogeneity. Due to symmetries of the stress and the strain tensors, only $21$ of these $81$ coefficients are independent. If the fault plane's material is assumed to be isotropic, the number of independent elastic coefficients further reduce to $2$. The Green's function term $(\mathbf{G}_{ij}(\mathbf{X},t;\pmb{\xi},\tau))$ describes the displacement in direction $i$ at a location $\mathbf{X}$ on the earth's surface in response to a concentrated impulse force in direction $j$ applied at $\pmb{\xi}$ on the fault plane (see Figure \ref{fig:schem-pwavedens_1}). The fault's time variable $\tau$ keeps a track of when this impulse force is applied while the ground surface time variable $t$ tracks how the ground displacement varies. This Green's function term: (1) is the earth's response to a unit point force (concentrated in space as well as time) applied on the fault plane; (2) propagates changes in displacement on the fault plane to those at the ground surface; (3) maintains appropriate time delays between displacements at the source and the site. The ground displacement of a scaled accelerogram can be further expressed as:

\begin{equation}
\begin{aligned}
U_i^\lambda(\mathbf{X},t) = \int_{\tau}d\tau \iint_{\Sigma} &\lambda_1 \big[U_j(\pmb{\xi},\tau)\big]~\lambda_2 c_{jkpq}~\lambda_3 \mathbf{G}_{ip,q}(\mathbf{X},t;\pmb{\xi},\tau)~\lambda_4 \nu_{k}~d\Sigma(\pmb{\xi})\\
& \textnormal{and}~\prod_{k = 1}^{4}\lambda_k = \lambda
\end{aligned}
\label{eqn:FactorSplit}
\end{equation}

\noindent
where the scale factor $\lambda$ has been split as sub-factors $\lambda_1,\lambda_2,\lambda_3,\textnormal{and}~\lambda_4$ which influence the terms $\big[U_j(\pmb{\xi},\tau)\big]$, $c_{jkpq}$, $\mathbf{G}_{ij}(\mathbf{X},t;\pmb{\xi},\tau)$\footnote{Scaling the first derivative of the Green's function by $\lambda$ also scales the Green's function by $\lambda$.}, and $\nu_{k}$, respectively, and which are constants with respect to space and time. It is noted that these sub-factors should satisfy the product condition provided in the second line of the above equation. In addition, if a sub-factor $\lambda_k$ does not influence its corresponding term in the representation theorem, its value will be unity. A physical interpretation on the behavior of each of these four sub-factors upon scaling a recorded accelerogram (i.e. when $\lambda > 1$ or $\lambda < 1$) is next made. In order to initiate this investigation, it is postulated that both the unscaled and the scaled accelerograms originate from the same seismic source; in other words, these accelerograms represent two separate seismic activities at the same fault plane. A scaled accelerogram represents a potential earthquake event that is yet to be realized. The above postulate attributes a seismic source to this unrealized earthquake event and thus provides a starting point for analyzing the seismology of a scaled accelerogram.  





\noindent
\subsection{Fault geometry and fault strength terms are invariant to accelerogram scaling}

The terms $\nu_k$ and $c_{jkpq}$ in equation \eqref{eqn:FactorSplit} establish a physical presence of the fault plane by attributing geometry and strength, respectively. The same seismic source postulate will be used discuss the scaling of these two terms.

\noindent
Term $\nu_k$ in equation \eqref{eqn:FactorSplit}, representing the unit normal vector of the fault plane and depicting the fault's geometry, cannot be freely scaled. As seen from Figure \ref{fig:schem-pwavedens_1}, for an example fault plane, the components of $\nu_k$ are $\{0.4,0.25,0.8818\}$\footnote{In a right-handed coordinate system.} satisfying the condition that the vector magnitude $\sqrt{\nu_k \nu_k} = 1$. All these components cannot arbitrarily be multiplied by some sub-factor $\lambda_4$ as such an operation will violate the definition of unit vector; consequently, $\lambda_4$ has to be unity and $\nu_k$ cannot be scaled. Additionally, the same seismic source postulate supports this argument because $\nu_k$, being a physical property of an existing fault, should be invariant to scaling.    

The term $c_{jkpq}$ in equation \eqref{eqn:FactorSplit}, depicting the elasticity of fault, converts displacements into forces. As per the same seismic source postulate, $c_{jkpq}$, being a physical property of an existing fault should again be invariant to scaling. Hence, the sub-factor $\lambda_2$ in equation \eqref{eqn:FactorSplit} must be unity. Anyhow, in order to facilitate further discussion on the same seismic source postulate, some mathematical implications concerning the elasticity coefficients and wave propagation velocities when $\lambda_2\neq 1$ are presented. Assuming isotropy of the fault's material, $c_{jkpq}$, when multiplied by $\lambda_2$, implies the following relation \citep{Aki2002}:

\begin{equation}
 \lambda_2~c_{jkpq} = (\lambda_2L_1)\delta_{jk} \delta_{pq} + (\lambda_2L_2) (\delta_{jp} \delta_{kq}+\delta_{jq} \delta_{kp})
\label{eqn:Elas1}
\end{equation}

\noindent
where $L_1$ and $L_2$ are the Lam\'e constants and $\delta_{xy}$ is a Kronecker delta which is equal to one when $x=y$ and zero otherwise. The above equation states that when the elasticity tensor $c_{jkpq}$ is multiplied by $\lambda_2$, both the Lam\'e constants are to be multiplied by the same sub-factor as the Kronecker delta, by definition, is either zero or one. As the Young's, Rigidity, and Bulk moduli ($E,~G,~K$, respectively) can be expressed using the two Lam\'e constants, it can be further shown that these elastic moduli are also multiplied by $\lambda_2$, while the Poisson's ratio ($\nu$) is held constant. Furthermore, it can be shown that material P-wave and S-wave velocities ($\alpha,~\beta$, respectively) are multiplied by $\sqrt{\lambda_2}$ while the material density is held constant. For example, following is the relation between scaled Lam\'e constants, P-wave velocity, and density $(\rho)$: 

\begin{equation}
\sqrt{\frac{\lambda_2~L_1+2\lambda_2~L_2}{\rho}} = \sqrt{\lambda_2}~\alpha
\label{eqn:Elas2}
\end{equation}

\noindent These mathematical implications suggest that if $\lambda_2 \neq 1$, there exists another seismic source which has the same ($\nu_k$, $\rho$, $\nu$)\footnote{Vector $\nu_k$ is the unit normal of a fault plane, whereas scalar $\nu$ is the Poisson's ratio.}, and spatial distribution of the elastic properties as the seismic source that generated the unscaled accelerogram, but ($E,~G,~K,~\alpha,$ and $\beta$) are modified by ${\lambda_2}$. Since it is intractable to ascertain the physical existence of this alternative seismic source for every arbitrary value of $\lambda_2$, the same seismic source postulate has been made. This postulate, while constraining $c_{jkpq}$ to be the same for both unscaled and scaled accelerograms, facilitates an investigation of the scaling of the other terms in the Representation theorem [equation \eqref{eqn:FactorSplit}].     

\subsection{The Green's function term: scaled and unscaled accelerograms are recorded at the same site}

\begin{figure}[h]
\begin{subfigure}{0.5\textwidth}
\centering
\includegraphics[width=0.88\textwidth]{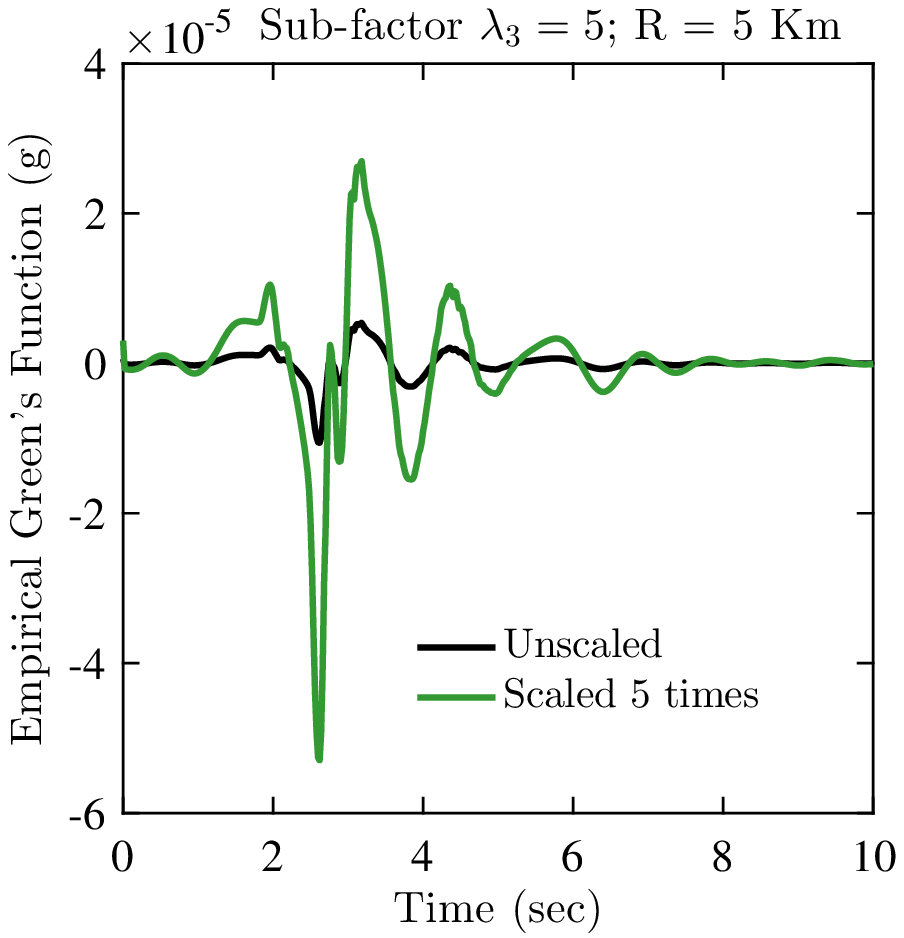}
\caption{} \label{fig:Geog_Greens}
\end{subfigure}\hspace*{\fill}
\begin{subfigure}{0.5\textwidth}
\centering
\includegraphics[width=0.88\textwidth]{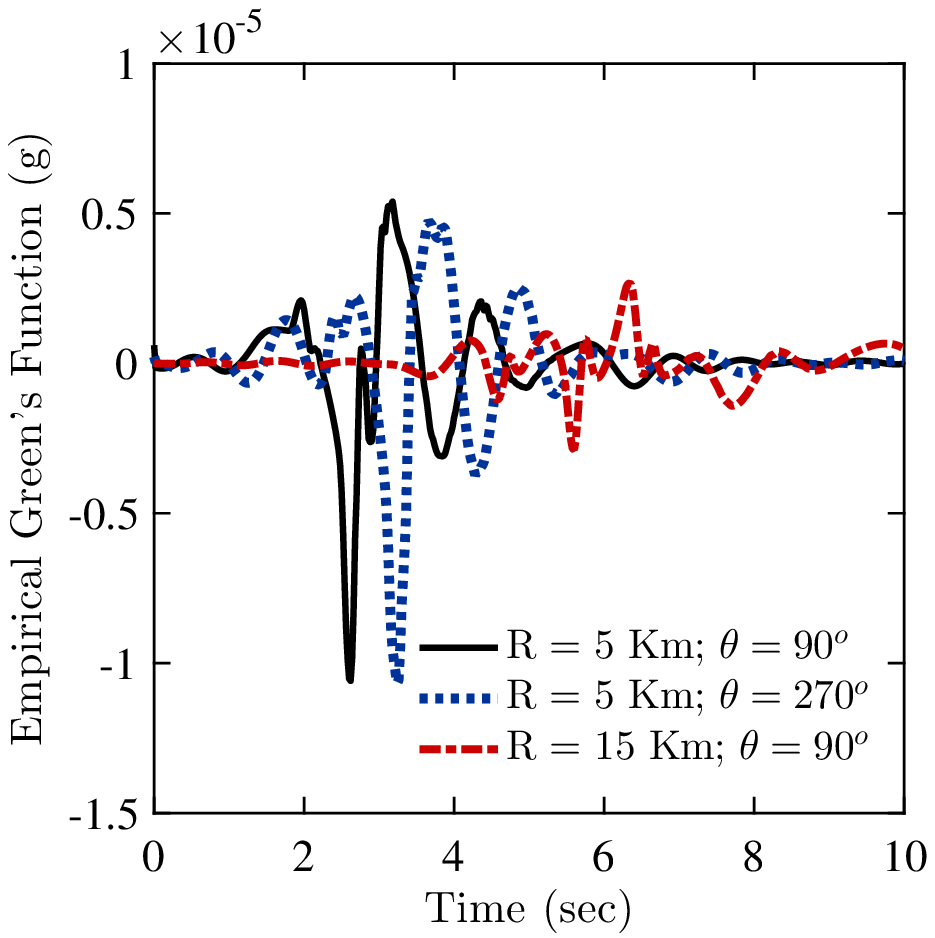}
\caption{} \label{fig:Dist_Greens}
\end{subfigure}
\caption{(a) Depiction of Green's function scaling using a simulated Empirical Green's function (EGF; \cite{Irikura1986}). This EGF is generated by a magnitude 3.0 earthquake using the SCEC BBP tool \citep{Maechling2015}. (b) EGFs for a magnitude 3 earthquake at three different sites. Angle $\theta$ here is the orientation of the recording station with respect to a horizontal through the epicenter.}\label{fig:EGF}
\end{figure}

The Green's function term $\big(\mathbf{G}_{ij}(\mathbf{X},t;\pmb{\xi},\tau)\big)$ in equation \eqref{eqn:FactorSplit} propagates displacements on the fault plane to those at the earth's surface and serves as a link between the source and the receiver. Figure \ref{fig:Geog_Greens} presents a pictorial depiction of $\mathbf{G}_{ij}(\mathbf{X},t;\pmb{\xi},\tau)$ scaling using Empirical Green's functions\footnote{Empirical Green's Functions are ground motions generated from small earthquakes whose sources may be characterized as impulsive point sources. An EGF thus approximates a component in the Green's function $\mathbf{G}_{ij}(\mathbf{X},t;\pmb{\xi},\tau)$. It should be noted that EGFs of different sites used here not only have the fault rupture (i.e., applied force) in the same direction but also the resulting motions are recorded in the same direction as well. In other words, the same component in $\mathbf{G}_{ij}(\mathbf{X},t;\pmb{\xi},\tau)$ for different sites will be compared.} (EGF; \cite{Irikura1986}) or approximate Green's functions. It is noted from this figure that both the unscaled and the scaled Green's functions have the same temporal shape in terms of start and end times, and polarities given any time. The only difference is, the scaled function has amplitudes greater by sub-factor $\lambda_3$ at all the times. $\mathbf{G}_{ij}(\mathbf{X},t;\pmb{\xi},\tau)$ scaling will be investigated using the Uniqueness theorem described below.

\noindent \textbf{Theorem A (Uniqueness):} \textit{Given an initial force distribution over the fault plane, the displacement at a site $\mathbf{X}$ is unique. In other words, given some initial conditions on the fault plane and $\mathbf{X}$, a single solution to the Representation theorem can exist.}

\noindent \textbf{Proof:} Interested readers may refer to Chapter 2, Section 2.3.1 in \cite{Aki2002} for a mathematical proof of Theorem A.
\QEDA


\noindent \textbf{Corollary A1:} \textit{$\mathbf{G}_{ij}(\mathbf{X},t;\pmb{\xi},\tau)$ in the Representation theorem is unique.} 

\noindent \textbf{Proof:} Given that a solution to the Representation theorem is unique, it is sufficient to show that $\mathbf{G}_{ij}(\mathbf{X},t;\pmb{\xi},\tau)$ is also a solution to the Representation theorem. For an applied force on the fault plane, the Representation theorem takes the form \citep{Aki2002}:

\begin{equation}
    \label{eqn:corr1}
    U_i(\mathbf{X},t) = \int_{\tau}d\tau \iiint_{V} f_j(\pmb{\xi},\tau)~\mathbf{G}_{ij}(\mathbf{X},t;\pmb{\xi},\tau)~dV(\pmb{\xi})
\end{equation}

\noindent where $f_j(\pmb{\xi},\tau)$ is force per unit volume. By definition, Green's function is the displacement at $\mathbf{X}$ for a unit impulse force applied at some point $\pmb{\xi}_1$ in space and $\tau_1$ in time. $f_j(\pmb{\xi},\tau)$ therefore is $\delta(\pmb{\xi}-\pmb{\xi}_1)~\delta(\tau-\tau_1)$ applied in direction $j$. Consequently:


\begin{equation}
    \label{eqn:corr1_1}
    U_i(\mathbf{X},t) = \int_{\tau}d\tau \iiint_{V} \delta(\pmb{\xi}-\pmb{\xi}_1)~\delta(\tau-\tau_1)~\delta_{ij}~\mathbf{G}_{ij}(\mathbf{X},t;\pmb{\xi},\tau)~dV(\pmb{\xi}) = \mathbf{G}_{ij}(\mathbf{X},t;\pmb{\xi}_1,\tau_1)
\end{equation}
\QEDA

Two additional corollaries that assign a spatial location to the scaled accelerogram are presented below. To discuss these corollaries, the inequivalence condition needs to be first introduced.

\textbf{Inequivalence condition:} Mathematically, this condition is given by $\mathbf{G}_{ij}(\mathbf{X},t;\pmb{\xi},\tau) \neq \mathbf{G}_{ij}(\mathbf{X}_1,t;\pmb{\xi},\tau)$ $\forall$ $\mathbf{X}_1$ $(\mathbf{X}_1 \neq \mathbf{X})$. The inequivalence condition suggests that given two different sites $\mathbf{X}$ and $\mathbf{X}_1$, not all components of their respective Green's function tensors $\big(\mathbf{G}_{ij}(\mathbf{X},t;\pmb{\xi},\tau)$ and $\mathbf{G}_{ij}(\mathbf{X}_1,t;\pmb{\xi},\tau)\big)$ can be the same. The reason is, two different sites will have different orientations (direction cosines) and/or distances with respect to the point of application of the unit impulse force $\pmb{\xi}$. This results in $\mathbf{G}_{ij}(\mathbf{X},t;\pmb{\xi},\tau)$ and $\mathbf{G}_{ij}(\mathbf{X}_1,t;\pmb{\xi},\tau)$ having either different polarities and/or durations. Figure \ref{fig:Dist_Greens} demonstrates these differences in Green's functions for three sites with different orientations and distances with respect to $\pmb{\xi}$. The solid black and dotted blue plots in Figure \ref{fig:Dist_Greens} demonstrate that for the same distance but different orientations of sites with respect to $\pmb{\xi}$, the corresponding Green's functions have dissimilar polarities. Alternatively, the solid black and dashed red plots in Figure \ref{fig:Dist_Greens} demonstrate that for the same orientation but different distances of sites with respect to $\pmb{\xi}$, the corresponding Green's functions have dissimilar start and end times, and hence durations.

\noindent \textbf{Corollary A2:} \textit{Given the uniqueness theorem and the inequivalence condition, the only value of $\lambda_3$ that is physically admissible in the operation $\lambda_3 \mathbf{G}_{ij}(\mathbf{X},t;\pmb{\xi},\tau)$ is unity.} 

\noindent \textbf{Proof:} Let us assume that $\lambda_3$ is anything but unity. Corollary A1 then prohibits attributing this scaled Green's function $\big(\lambda_3 \mathbf{G}_{ij}(\mathbf{X},t;\pmb{\xi},\tau)\big)$ to a site with some value of $\mathbf{X}_1$. This is because, given $\mathbf{X}_1$, its Green's function $\mathbf{G}_{ij}(\mathbf{X}_1,t;\pmb{\xi},\tau)$ is unique. The $\lambda_3 \mathbf{G}_{ij}(\mathbf{X},t;\pmb{\xi},\tau)$ additionally cannot equal $\mathbf{G}_{ij}(\mathbf{X}_1,t;\pmb{\xi},\tau)$ because the inequivalence condition suggests that the elements of these Green's functions will have different polarities and/or durations. Therefore, $\lambda_3$ cannot be anything but unity.
\QEDA

\noindent \textbf{Corollary A3:} \textit{Given the inequivalence condition, both unscaled and scaled accelerograms correspond to the same location $\mathbf{X}$.} 

\noindent \textbf{Proof:} By constraining $\lambda_3$ value to unity in equation \eqref{eqn:FactorSplit}, Corollary A2 implies that both unscaled and scaled accelerograms have the same Green's functions $\big(\mathbf{G}_{ij}(\mathbf{X},t;\pmb{\xi},\tau)\big)$. Additionally, since $\mathbf{G}_{ij}(\mathbf{X},t;\pmb{\xi},\tau) \neq \mathbf{G}_{ij}(\mathbf{X}_1,t;\pmb{\xi},\tau)$ under $\mathbf{X}\neq \mathbf{X}_1$, the only location $\mathbf{G}_{ij}(\mathbf{X},t;\pmb{\xi},\tau)$ can correspond to for the scaled accelerogram is $\mathbf{X}$.
\QEDA




\subsection{The rupture term: magnitude and distance values for a scaled accelerogram}

The term $\big[U_j(\pmb{\xi},\tau)\big]$ in equation \eqref{eqn:FactorSplit} represents rupture on the fault plane and has a dimension of a length. With the other three sub-factors in equation \eqref{eqn:FactorSplit} ($\lambda_2$, $\lambda_3$, and $\lambda_4$) equal to unity, if a recorded motion is to be scaled by $\lambda$, then, the sub-factor corresponding to $\big[U_j(\pmb{\xi},\tau)\big]$ ($\lambda_1$) should be equal to $\lambda$. In other words, scaling of the recorded motion must be attributed to scaling of the rupture amplitudes on the fault plane. The degree to which this attribution can be made depends on the dynamics of the earthquake source process. Source dynamics is a complicated topic that requires an understanding of advanced mathematics along with fracture mechanics. So, an empirical approach is used to discuss $\big[U_j(\pmb{\xi},\tau)\big]$ scaling. First, this term's scaling is mapped here to seismic parameters of engineering interest such as rupture extent, distance, and magnitude. Later on in this paper, empirical relationships will be discussed to constrain the scaling of these seismic parameters, which also implicitly constrains $\big[U_j(\pmb{\xi},\tau)\big]$ scaling.



\begin{figure}[h]
\begin{subfigure}{0.5\textwidth}
\centering
\includegraphics[width=0.85\textwidth]{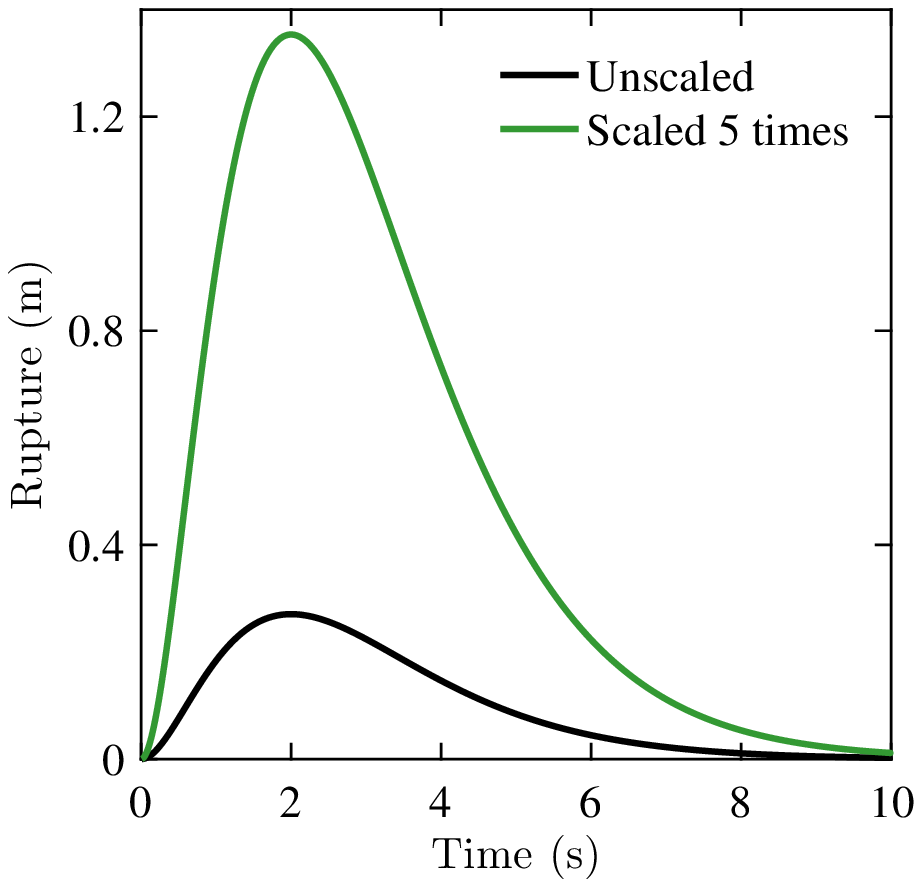}
\caption{} \label{fig:STF}
\end{subfigure}\hspace*{\fill}
\begin{subfigure}{0.5\textwidth}
\centering
\includegraphics[width=0.85\textwidth]{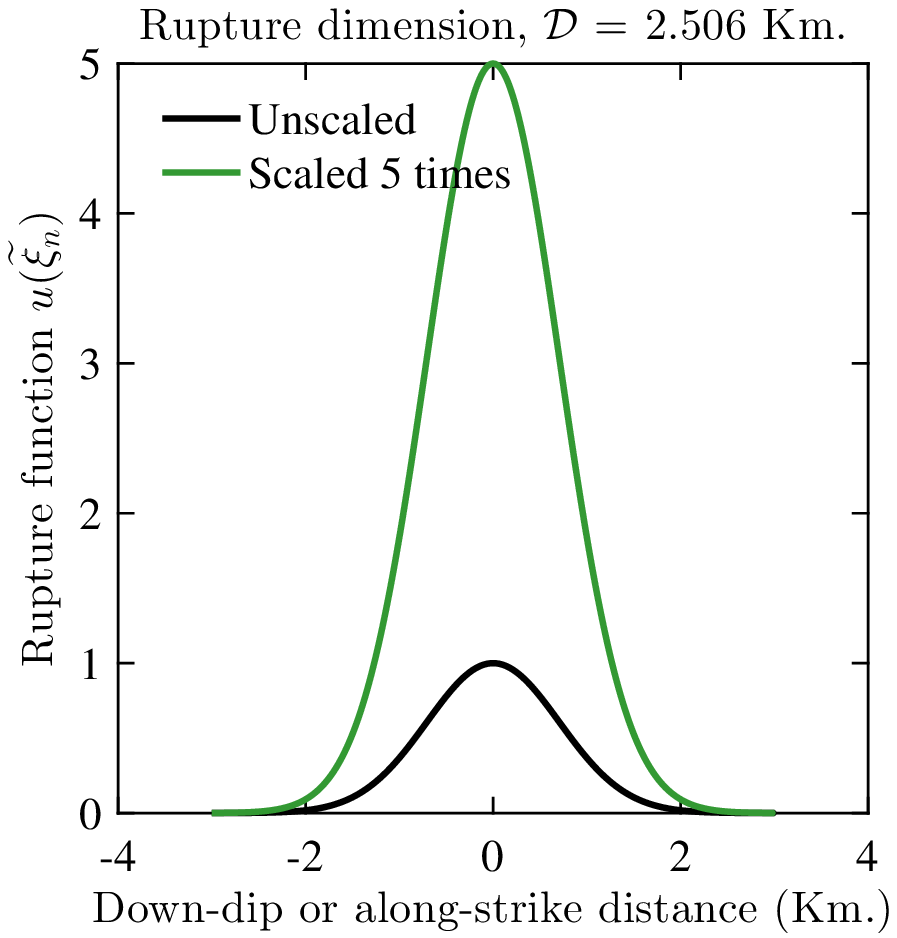}
\caption{} \label{fig:autocorr}
\end{subfigure}
\caption{(a) Illustration of the rupture dependence on time for unscaled and scaled rupture functions using a Gamma like function. It is observed that both the functions depend on time in the same manner. (b) Illustration of the rupture dependence on space for unscaled and scaled rupture functions using a Gaussian function for rupture distribution. It is observed that both the functions have the same spatial distribution of rupture. As a result, the effective rupture dimension computed using the \cite{Mai2000} definition $(\mathcal{D} = 2.506$ Km) is the same for both the unscaled and the scaled functions.} \label{fig:source}
\end{figure}

From a kinematic viewpoint, multiplying $\lambda$ to $\big[U_j(\pmb{\xi},\tau)\big]$ increases the amplitudes of the rupture without influencing either the spatial distribution of rupture (rupture extent) or the dependence of rupture on time. Figure \ref{fig:STF} presents an illustration of the rupture dependence on time for unscaled and scaled rupture functions using a Gamma like function. It is observed that both the functions depend on time in the same manner. 



Invariance of the rupture extent to accelerogram scaling has implications on the effective rupture dimensions. Consider the one-dimensional rupture function $[{u}(\widetilde{\xi}_n)]$ computed by summing the time-averaged rupture values across the down-dip (or along-strike) direction (\cite{Mai2000})\footnote{The notation $\widetilde{\pmb{\xi}}$ indicates a fault plane coordinate system. In this system, while two directions are along-strike and down-dip of the fault plane, the third direction along which rupture amplitudes do not vary is normal to the fault plane. The subscript $n$ in $[{u}(\widetilde{\xi}_n)]$ therefore denotes either the down-dip or along-strike direction. Additionally, the fault plane system can be different from the general coordinate system $\pmb{\xi}$ to represent a point on the fault plane.}. Linearly scaling $\big[U_j(\pmb{\xi},\tau)\big]$ also scales $[{u}(\widetilde{\xi}_n)]$ linearly because the coordinate transformation from $\pmb{\xi}$ to $\pmb{\widetilde{\xi}}$ is linear; in other words, $[u^\lambda(\widetilde{\xi}_n)] = \lambda~[u(\widetilde{\xi}_n)]$. The following theorem demonstrates an implication of linearly scaling $[u(\widetilde{\xi}_n)]$ on the effective rupture dimension $(\mathcal{D})$. A corollary then shows the invariance of the Joyner-Boore distance metric to accelerogram scaling.




\noindent \textbf{Theorem B:} \textit{Provided that the effective rupture dimension $(\mathcal{D})$ is defined as per \cite{Mai2000}, $\mathcal{D}$ is invariant to accelerogram scaling.}

\noindent \textbf{Proof:} Given that $[u^\lambda(\widetilde{\xi}_n)] = \lambda~[u(\widetilde{\xi}_n)]$, it is straightforward to show that the \cite{Mai2000} definition for $\mathcal{D}$ is invariant to accelerogram scaling since the scale factors in the numerator and the denominator cancel each other out:

\begin{equation}
    \label{eqn:rupDim}
    \mathcal{D} = \frac{\int_{\mathcal{L}}~d\mathcal{L}\int_{\widetilde{\xi}_n}~[u(\widetilde{\xi}_n)]~[u(\widetilde{\xi}_n-\mathcal{L})]~d\widetilde{\xi}_n}{\int_{\widetilde{\xi}_n}~[u(\widetilde{\xi}_n)]~[u(\widetilde{\xi}_n)]~d\widetilde{\xi}_n} = \frac{\int_{\mathcal{L}}~d\mathcal{L}\int_{\widetilde{\xi}_n}~[u^\lambda(\widetilde{\xi}_n)]~[u^\lambda(\widetilde{\xi}_n-\mathcal{L})]~d\widetilde{\xi}_n}{\int_{\widetilde{\xi}_n}~[u^\lambda(\widetilde{\xi}_n)]~[u^\lambda(\widetilde{\xi}_n)]~d\widetilde{\xi}_n}
\end{equation}

\noindent where $\mathcal{L}$ is the lag-length in an autocorrelation function.
\QEDA

\noindent \textbf{Corollary B:} \textit{The Joyner-Boore distance metric $(R_{JB})$ is invariant to accelerogram scaling.}

\noindent \textbf{Proof:} $R_{JB}$ is dependent on two quantities: the location of a site $\mathbf{X}$ and the surface projection of the effective rupture dimension $\mathcal{D}$. Corollary A3 suggests that both unscaled and scaled accelerograms are recorded at the same site $\mathbf{X}$. Theorem B suggests that both unscaled and scaled accelerograms have the same effective rupture dimension $\mathcal{D}$, and therefore its surface projection. As a result of this invariance of the two quantities upon which $R_{JB}$ depends upon, $R_{JB}$ is also invariant to accelerogram scaling.
\QEDA

Figure \ref{fig:autocorr} presents an illustration of the constancy of $\mathcal{D}$ to accelerogram scaling by assuming a Gaussian function for rupture. Coming to earthquake strength, the following theorem and corollaries are proposed.

\noindent \textbf{Theorem C:} \textit{The seismic moment tensor $(\mathbf{M}_{pq})$ scales linearly with accelerogram scaling.}

\noindent \textbf{Proof:} The proof follows from the definition of $\mathbf{M}_{pq}$ \citep{Aki2002}\footnote{By averaging over time, the time dependence of the moment tensor has been excluded.}:

\begin{equation}
    \label{eqn:mom_ten}
    \mathbf{M}_{pq}^{\lambda} = \frac{\int_{\tau} d\tau \iint_{\Sigma} \lambda \big[U_i(\pmb{\xi},\tau)\big]~\nu_j~c_{ijpq}~d\Sigma(\pmb{\xi})}{\int_{\tau} d\tau} = \lambda~\mathbf{M}_{pq}
\end{equation}

\noindent where $\mathbf{M}_{pq}^{\lambda}$ is the moment tensor for the scaled accelerogram. It is observed from the above equation that linearly scaling the rupture amplitudes scales the moment tensor linearly.
\QEDA

\noindent \textbf{Corollary C1:} \textit{Seismic moment $(M_o)$ scales linearly with accelerogram scaling.}

\noindent \textbf{Proof:} The proof follows from equation \eqref{eqn:mom_ten} and the definition of $M_o$ \citep{Shearer2009}:

\begin{equation}
    \label{eqn:moment}
    M_o^{\lambda} = \frac{1}{\sqrt{2}}~\big(\mathbf{M}_{pq}^{\lambda}\mathbf{M}_{pq}^{\lambda}\big)^{\frac{1}{2}} = \lambda~M_o
\end{equation}

\noindent where ${M}_{o}^{\lambda}$ is the seismic moment for the scaled accelerogram. It is observed from the above equation that linearly scaling the moment tensor scales the seismic moment linearly.
\QEDA

\noindent \textbf{Corollary C2:} \textit{Moment magnitude $(M_w)$ scales logarithmically with accelerogram scaling.}

\noindent \textbf{Proof:} The proof follows from equation \eqref{eqn:moment} and the definition of $M_w$ \citep{Hanks1979}:

\begin{equation}
    \label{eqn:magn}
    M_w^{\lambda} = \frac{2}{3}~log_{10}\big(M_o^{\lambda}\big)-6.07 = M_w + \frac{2}{3}~log_{10}(\lambda)
\end{equation}

\noindent where ${M}_{w}^{\lambda}$ is the moment magnitude for the scaled accelerogram. It is observed from the above equation that linearly scaling the moment tensor scales the moment magnitude logarithmically.
\QEDA

\section{Insights from Fourier Amplitude spectrum of an accelerogram}\label{Fourier}

Previously, the Representation theorem was used to gain insights into ground motion scaling in the time domain. Whereas magnitudes of the unscaled and scaled accelerograms were shown to be related by equation \eqref{eqn:magn}, it was concluded that effective rupture dimensions, location of the recording station, and hence the Joyner-Boore distance metric remain unchanged upon scaling an accelerogram. In this section, accelerogram scaling will be investigated in the frequency domain so as to map such scaling to parameters such as the Brune's static stress drop and the corner frequency. The additional insights that are gained in this section will be in reconciliation with those obtained from the Representation theorem. 

\begin{figure}[h]
\begin{subfigure}{0.5\textwidth}
\centering
\includegraphics[width=0.85\textwidth]{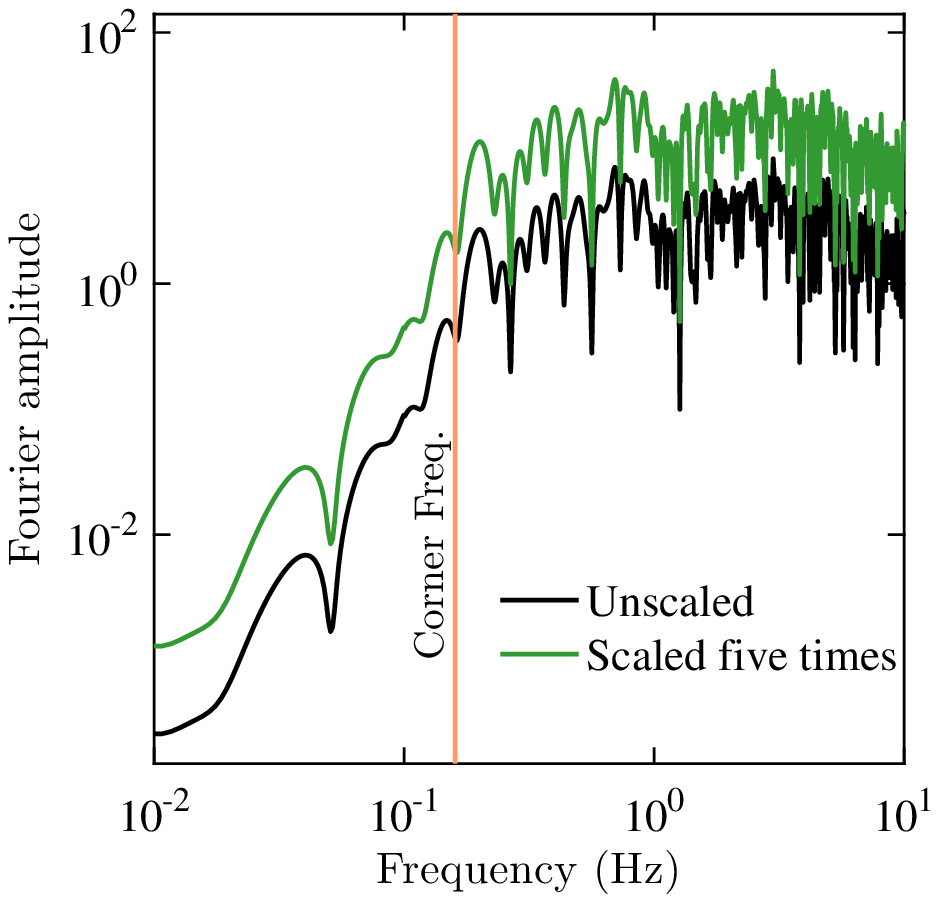}
\caption{} \label{fig:scaled_freq}
\end{subfigure}\hspace*{\fill}
\begin{subfigure}{0.5\textwidth}
\centering
\includegraphics[width=0.85\textwidth]{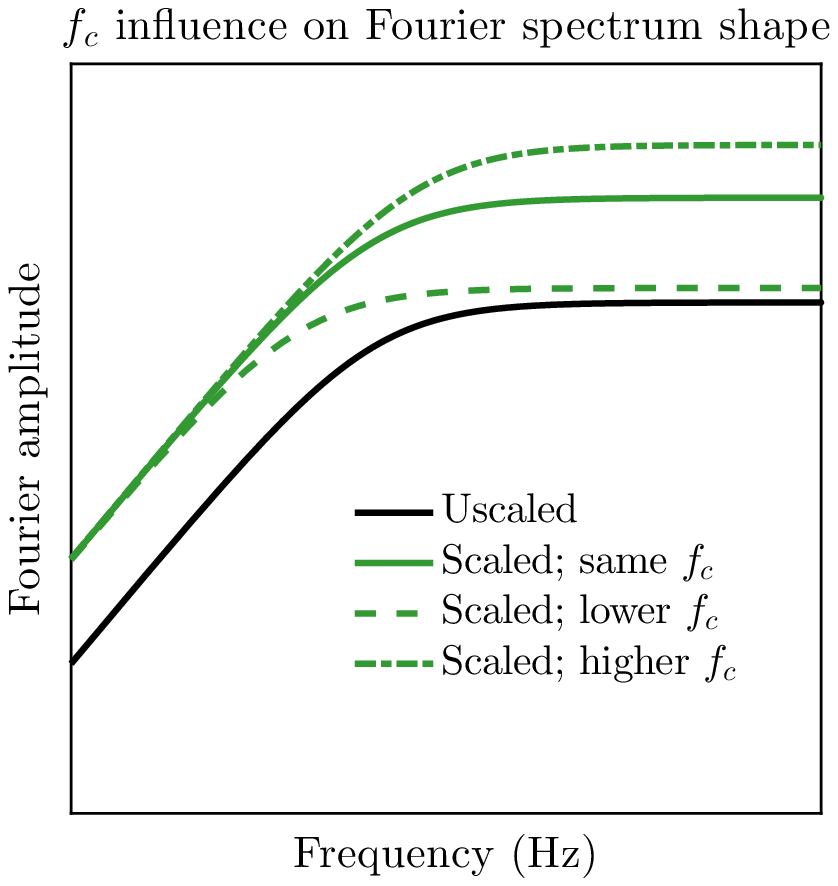}
\caption{} \label{fig:model_freq}
\end{subfigure}
\caption{(a) Fourier amplitude spectra of an unscaled and a scaled accelerogram from the Northridge earthquake. The shape of the spectra are observed to be the same with a corner frequency $(f_c)$ of $0.16$Hz. (b) Illustration of the influence of $f_c$ on the Fourier spectrum shape. Only by keeping $f_c$ the same for unscaled and scaled spectra, we obtain a scaled Fourier spectrum that is uniformly scaled at all frequencies.}\label{fig:freq}
\end{figure}

Figure \ref{fig:scaled_freq} presents Fourier amplitude spectra of unscaled and scaled accelerograms from the Northridge earthquake. It can be noticed that accelerogram scaling influences neither the shape nor the frequency content of the Fourier amplitude spectrum. The amplitudes corresponding to the different frequencies, however, are scaled uniformly by the scale factor $\lambda$. As an aside, an overarching insight provided by the Representation theorem is: ground motion scaling influences the source kinematics in terms of amplifying the rupture amplitudes by $\lambda$ without changing neither the Green's function (the propagator, accounts for path effects) nor the source geometric and materialistic characteristics. Combining this insight with the observations made regarding the shape of the Fourier spectra in Figure \ref{fig:scaled_freq}, the following expression models the Fourier spectrum of a scaled accelerogram \citep{Boore2003}:

\begin{equation}
    \label{eqn:FAS}
    \pmb{\mathcal{F}}^{\lambda}(f, M^\lambda_o, R) = I(f)~P(R, f)~E^{\lambda}(M^\lambda_o, f)~G(f)
\end{equation}

\noindent where $I(f)$, $P(R,f)$, $E^{\lambda}(M^\lambda_o, f)$, and $G(f)$ are the ground motion type, path, scaled source terms, and site-response, respectively. $I(f)$ is invariant to accelerogram scaling as it only accounts for whether the resulting ground motion is acceleration, velocity, or displacement. The scaling of the other three terms is discussed subsequently.

\subsection{The path term is invariant to accelerogram scaling}

The path term $P(R,f)$ in the model for Fourier spectrum of an accelerogram [equation \eqref{eqn:FAS}] connects seismic activity on the fault plane to displacements at a site. This term's purpose is qualitatively similar to that of the Green's function in the Representation theorem. Since the Green's function was shown to remain invariant to accelerogram scaling, one may speculate that this invariance holds for $P(R,f)$ as well. As shown below, such a speculation is supported by the manner in which $P(R,f)$ is usually defined \citep{Boore2003}:

\begin{equation}
    \label{eqn:Path}
    P(R,f) = e^{\frac{-\pi f R}{Q(f)\beta}}~Z(R)
\end{equation}

\noindent where $R$ is a distance metric between the source and the receiver, $\beta$ is the shear-wave velocity in the source region, and $Q(f)$ is the quality factor which accounts for inelastic attenuation of the seismic waves. $Z(R)$ in the above equation accounts for geometric attenuation and is proportional to $R^{-1}$ in the simplest case of linear geometric attenuation. In summary, $P(R,f)$, apart from frequency, is seen to depend on the distance metric $R$, which is usually defined using the Joyner-Boore distance metric. Hence, $P(R,f)$ should be invariant to accelerogram scaling since $R_{JB}$ was noted to remain unchanged upon scaling an accelerogram.


\subsection{The scaled source term: Stress drop scales linearly with accelerogram scaling}

The scaled source term in equation \eqref{eqn:FAS} can be expanded using a source model. An $\omega$-squared source model \citep{Brune1970, Boore2003}, which characterizes the fault plane as a point source, is adopted here. The scaled source term is:

\begin{equation}
    \label{eqn:source}
    E^{\lambda}(M^\lambda_o, f) = \frac{\langle R_{\Theta \Phi} \rangle}{2\sqrt{2}\pi \rho_s \beta_s^3}~M_o^\lambda~\frac{f^2}{1+(\frac{f}{f_c})^2}
\end{equation}

\noindent where $\langle R_{\Theta \Phi} \rangle$ is the average radiation coefficient which depends on the geometry of the source; $\rho_s$ and $\beta_s$ are respectively the density and the shear-wave velocity of the material at the source. Because accelerogram scaling will not alter the source geometric and materialistic properties (\textit{Same Seismic Source} axiom), $\langle R_{\Theta \Phi} \rangle$, $\rho_s$, and $\beta_s$ will not change for unscaled and scaled accelerograms. In equation \eqref{eqn:source}, $M_o^\lambda$ is the scaled seismic moment given by equation \eqref{eqn:moment}; $f$ is the frequency of interest and $f_c$ is the corner frequency which distinguishes the low and high frequency parts of the Fourier spectrum. $f_c$ remains unchanged upon scaling an accelerogram. This is because, as per Figure \ref{fig:scaled_freq}, the shape of the Fourier spectrum should remain the same for both unscaled and scaled accelerograms. Any deviations in the $f_c$ value between unscaled and scaled accelerograms would result in differences in the spectral shapes and hence non-uniform scaling of the spectrum across the frequencies. Figure \ref{fig:model_freq} presents an illustration of the influence of $f_c$ on the Fourier spectrum shape. It is observed from this figure that, only by keeping $f_c$ for the scaled spectrum equal to that of the unscaled spectrum, we obtain a scaled Fourier spectrum that is uniformly scaled at all frequencies (which is in corroboration with Figure \ref{fig:scaled_freq}). 


Same $f_c$ for unscaled and scaled accelerograms has two important implications. First, $f_c$ is related to the effective rupture dimensions by some source models. For example, the Brune's and the Haskell's models relate $f_c$ to the radius and the length/width of the rupture, respectively \citep{Shearer2009}. Constancy of $f_c$ as per these models implies both unscaled and scaled accelerogram result from the same effective rupture dimensions. Insights from the Representation theorem also led to the same conclusion. Second, interpreting the Fourier spectrum of a scaled accelerogram using the Brune's source model, $f_c$ is related to $M_o^\lambda$ as \citep{Brune1970, Baltay2014}:

\begin{equation}
    \label{eqn:Brune}
    \begin{aligned}
    & f_c \propto \Big(\frac{\Delta \sigma^\lambda}{M_o^\lambda}\Big)^{\frac{1}{3}}\\
     \textnormal{and},~\Delta &\sigma^\lambda = \lambda~\Delta \sigma; M_o^\lambda = \lambda~M_o\\
    \end{aligned}
\end{equation}

\noindent where $M_o^\lambda$ is the scaled seismic moment and $\Delta \sigma^\lambda$ is the scaled Brune's stress drop. To ensure constancy of $f_c$ for both unscaled and scaled accelerograms, $\Delta \sigma^\lambda$ must be equal to $\lambda~\Delta \sigma$ as scaling an accelerogram scales the seismic moment linearly. That is, Brune's stress drop of the scaled accelerogram $(\Delta \sigma^\lambda)$ is scale factor $(\lambda)$ times the Brune's stress drop of the unscaled accelerogram $(\Delta \sigma)$. 

Synthesis of accelerogram scaling in the Fourier domain, which was made using a point seismic source until now, can be also extended to finite seismic sources. The scaled source term for a finite source can be expressed as \citep{Motazedian2005}\footnote{It is noted that there is an implicit summation over indices $ij$ in the expression for $E^\lambda(M_o^\lambda,f)$ in equation \eqref{eqn:sourceFin}.}:

\begin{equation}
    \label{eqn:sourceFin}
    \begin{aligned}
    E^\lambda(M_o^\lambda,f) &=  \frac{\langle R_{\Theta \Phi} \rangle}{2\sqrt{2}\pi \rho_s \beta_s^3}~M_{oij}^\lambda~\frac{f^2}{1+(\frac{f}{f_{cij}})^2}\\
    \textnormal{and,}~f_{cij} \propto \Big(\frac{\Delta \sigma^\lambda}{\overline{M}_{o}^\lambda}\Big)&^{\frac{1}{3}};~M_{oij}^\lambda = \lambda~M_o~\mathcal{W}_{ij};~\overline{M}_{o}^\lambda = \lambda~M_o/N\\
    \end{aligned}
\end{equation}

\noindent where $ij$ are the indices of a sub-fault among the $N$ sub-faults into which the seismic source is discretized and $\mathcal{W}_{ij}$, the slip weight of the $ij$ sub-fault, satisfies $\mathbf{1}_{ij}\mathcal{W}_{ij} = 1$. To ensure constancy of the sub-fault corner frequency $f_{cij}$ when $M_o$ is multiplied by $\lambda$, it is again seen that $\Delta \sigma^\lambda = \lambda~\Delta \sigma$. That is, even in the case of a finite seismic source, stress drop scales linearly with accelerogram scaling.

\subsection{The site term: Relative nonlinearity of site response must not be large}

Sites where earthquake ground motion is recorded are usually resting on a few tens of meters of soil layers that lie above bedrock (see Figure \ref{fig:schem-pwavedens_1}). Although seismic waves travel a large distance through rock and a relatively small distance through soil, the soil plays an important role in characterizing the surface ground motion \citep{Kramer1996}. The soil's role in influencing the surface ground motion is termed as site response, which is explicitly modeled using the $G(f)$ term in equation \eqref{eqn:FAS}. Site response is categorized as nonlinear or linear depending upon whether it is affected by the strength of the bedrock accelerogram or not. With respect to accelerogram scaling, it is important that the relative nonlinearity of site response is not very large.

\begin{figure}[h]
\begin{subfigure}{0.333\textwidth}
\centering
\includegraphics[width=1\textwidth]{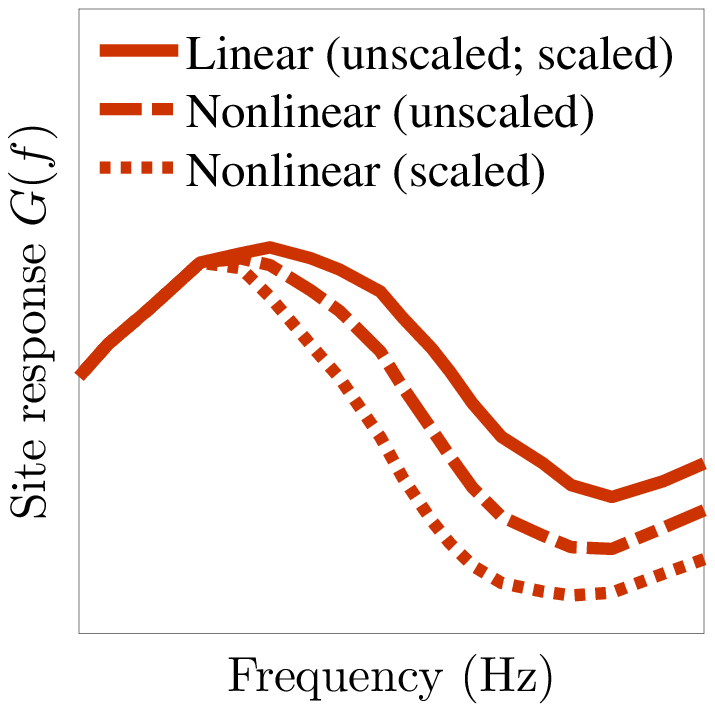}
\caption{} \label{fig:Site_schem}
\end{subfigure}\hspace*{\fill}
\begin{subfigure}{0.333\textwidth}
\centering
\includegraphics[width=1\textwidth]{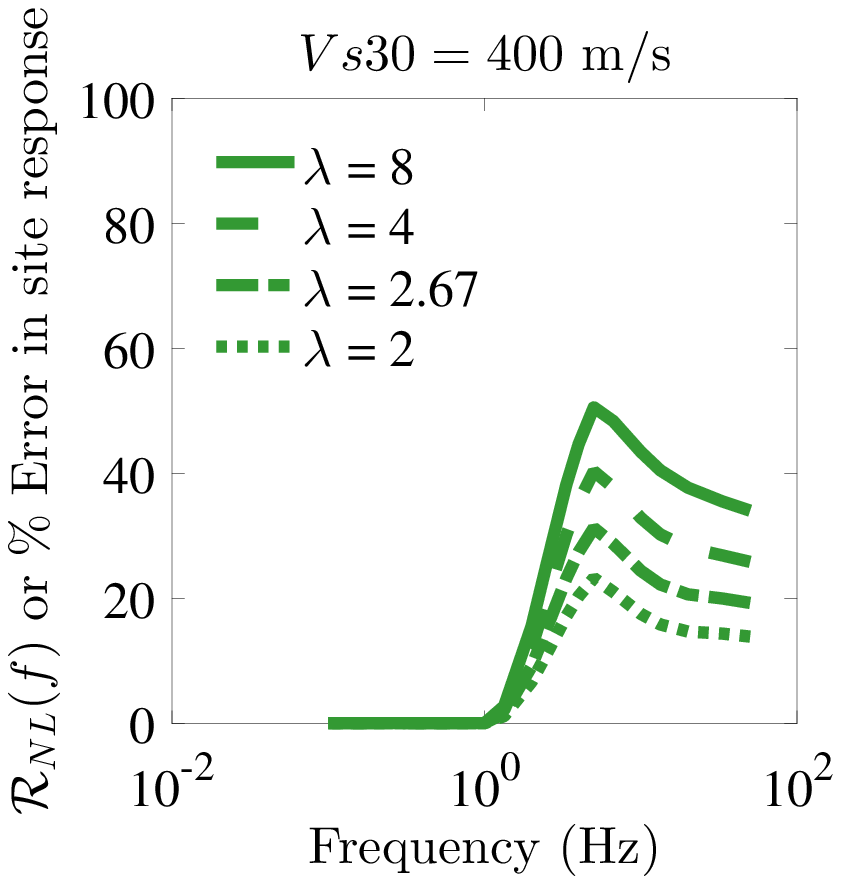}
\caption{} \label{fig:Site_400}
\end{subfigure}\hspace*{\fill}
\begin{subfigure}{0.333\textwidth}
\centering
\includegraphics[width=1\textwidth]{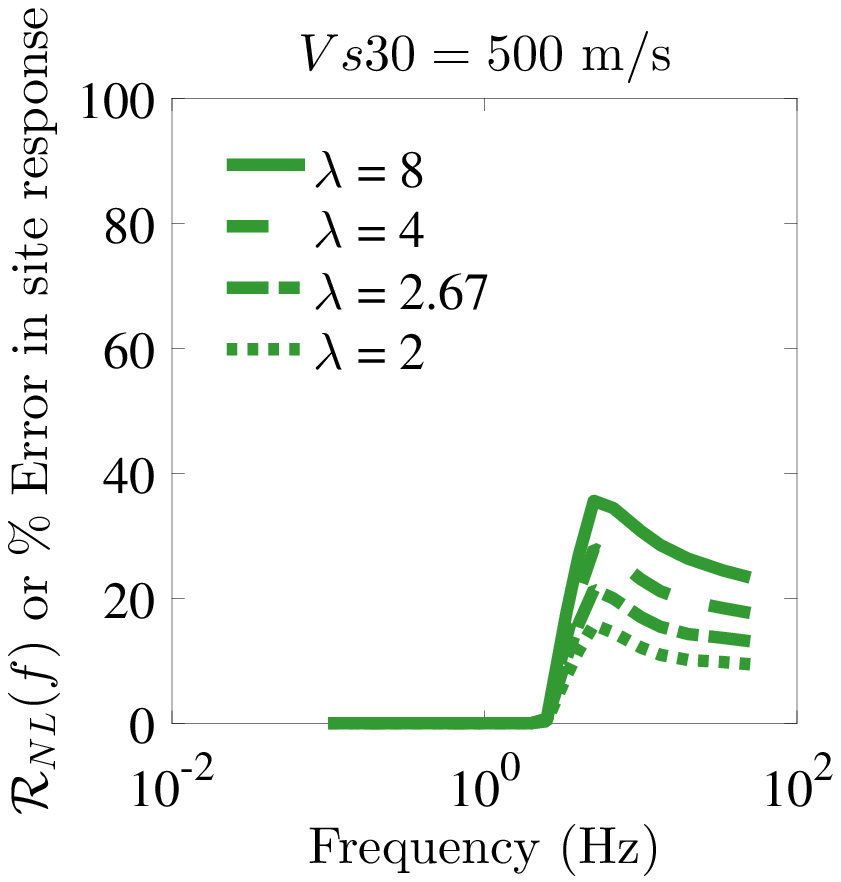}
\caption{} \label{fig:Site_500}
\end{subfigure}
\caption{(a) Schematic of linear and nonlinear site response functions $G(f)$. It is noted that nonlinear site response, due to its dependence on the strength of the bedrock motion, differs for unscaled and scaled bedrock motions. (b) and (c): Relative nonlinearity of site response $\mathcal{R}_{NL}(f)$ defined in terms of a percentage error; see equation \eqref{eqn:SiteErr}. $\mathcal{R}_{NL}(f)$ values are presented for two sites with $Vs30$ as (b) 400 $m/s$ and (c) 500 $m/s$ considering four levels of bedrock accelerogram scaling.}\label{fig:Site}
\end{figure}

Relative nonlinearity of site response is defined as how much more (or less) nonlinear the site response is for the scaled accelerogram as compared to the unscaled one. A scaled accelerogram can be generated by scaling the seismic parameters $M_w$ and $\Delta \sigma$ using scale factor $\lambda$ as discussed previously. However, this scaled accelerogram will not be $\lambda$ times the unscaled one should the relative nonlinearity of site response be large. For example, if the soil underlying a site is very dense, its site response will be close to linear and accelerograms generated by both unscaled and scaled seismic parameters will have the same linear site response function (the solid plot in Figure \ref{fig:Site_schem}). There is no relative nonlinearity of site response in this case. On the other hand, if the soil underlying a site is weak, its site response will be nonlinear and thus will differ for the unscaled and the scaled values of seismic parameters (the dashed and the dotted plots in Figure \ref{fig:Site_schem}, respectively). The reason for such difference is, the scaled seismic parameters (assuming scaled upwards) will produce a stronger bedrock motion as compared to the unscaled seismic parameters, and the nonlinear soil response cannot transmit both the intense and the less intense bedrock motions to the surface in the same manner.     

Mathematically, relative nonlinearity of site response $\mathcal{R}_{NL}(f)$ is defined in terms of a percentage error as:

\begin{equation}
    \label{eqn:SiteErr}
    \mathcal{R}_{NL}(f)~\textnormal{or}~\%~\textnormal{Error} = \frac{\mid G(f)-G^\lambda(f)\mid}{G(f)}\times 100
\end{equation}

\noindent where $G(f)$ and $G^\lambda(f)$ are the site response terms for the unscaled and the scaled accelerograms, respectively. These site response terms, in addition to frequency, also depend upon the quality of soil at a site and the strength of the bedrock motion. Selecting the site model $F_{site}(.)$ of the \cite{Campbell2014} ground motion prediction model, site response terms for the unscaled and the scaled seismic parameters can be expressed as\footnote{While the \cite{Campbell2014} site model strictly applies to response spectra, its use for Fourier spectra is justified because the model varies smoothly with frequency \citep{Graves2010a}.}:

\begin{equation}
    \label{eqn:Site_CB}
    G(f) = \frac{F_{site}(f,Vs30,PGA_{rock})}{F_{site}(f,Vs_{ref},PGA_{rock})};~~~~G^\lambda(f) = \frac{F_{site}(f,Vs30,PGA^\lambda_{rock})}{F_{site}(f,Vs_{ref},PGA^\lambda_{rock})}
\end{equation}

\noindent where $Vs30$ is the shear-wave velocity averaged over the top $30$ meters depth, $Vs_{ref}$ is the reference shear-wave velocity set to $760$ m/s, and $PGA_{rock}$ and $PGA^\lambda_{rock}$ are the Peak Ground Accelerations of the bedrock under unscaled and scaled values of the seismic parameters, respectively. It is noted that while $Vs30$ serves as a proxy for the quality of soil at a site, $PGA_{rock}$ and $PGA^\lambda_{rock}$ serve as proxies for describing the strength of the bedrock motion. Furthermore, $PGA^\lambda_{rock} = \lambda~PGA_{rock}$ because the scaled seismic parameters are expected to generate a bedrock accelerogram that is $\lambda$ times the bedrock accelerogram generated by the unscaled seismic parameters.

\noindent Figures \ref{fig:Site_400} and \ref{fig:Site_500} present relative nonlinearity of site response $\mathcal{R}_{NL}(f)$ for two sites with $Vs30$ values as $400~m/s$ and $500~m/s$, respectively. These values of $Vs30$ indicate that the former site has weaker soil as compared to latter site. Bedrock PGA generated by the scaled seismic parameters $(PGA_{rock}^\lambda)$ is set to $2g$, and four levels of unscaled bedrock PGA $(PGA_{rock})$ are considered: $0.25g$, $0.5g$, $0.75g$, and $1g$. In other words, $\mathcal{R}_{NL}(f)$ is computed by selecting scale factors $(\lambda)$ as 8, 4, 2.67, and 2. From both Figures \ref{fig:Site_400} and \ref{fig:Site_500}, it is observed that as the scale factor increases, $\mathcal{R}_{NL}(f)$ also increases. This is expected because higher scale factors result in more intense bedrock motions which in turn induce more nonlinearity in the soil. The relative nonlinearity of site response is also observed to be more profound for the weaker soil (Figure \ref{fig:Site_400}) as compared to the stiffer soil (Figure \ref{fig:Site_500}). Considering this joint role played by $\lambda$ and $Vs30$, it can be generally stated that accelerograms recorded on stiff soil (or better) sites facilitate accelerogram scaling to a greater degree. The reason is, such sites have a better capacity to linearly transmit intense bedrock motions than weak soil sites.       

\section{Validating the seismological interpretation of accelerogram scaling}

Accelerogram scaling can be described in terms of similarities and differences between certain seismological variables. Both unscaled and scaled accelerograms will have the same: spatial and temporal variation of rupture at the same seismic source, effective rupture dimensions (and hence the rupture area), Joyner-Boore distance metric, and corner frequency. Brune's stress drops and magnitudes, however, will be different. Stress drop of the scaled accelerogram is $\lambda$ times stress drop of the unscaled one, and magnitude of the scaled accelerogram is: $M_w^\lambda = M_w + 2/3~log_{10}(\lambda)$. If two accelerograms satisfy these seismological conditions under similar wave attenuation and site response effects, then in principle, one accelerogram should be $\lambda$ times the other.

\subsection{Use of ground motion simulations for validation}

In order to validate the proposed interpretation of scaling, we need to compare two accelerograms that satisfy the above-mentioned conditions. In practice, since it is difficult to find two recorded accelerograms that satisfy these conditions, reliance will be made upon ground motion simulations. First, an accelerogram corresponding to a particular seismic source and distance will be simulated with appropriate input values for $M_w$, $f_c$, and rupture dimensions. This will be treated as an unscaled accelerogram. Next, a second accelerogram with magnitude given by equation \eqref{eqn:magn} and all other inputs same as the previous run will be simulated. This will be treated as a scaled simulated accelerogram. Finally, a comparison will be made between these two accelerograms after explicitly scaling the first accelerogram by factor $\lambda$ (termed as explicitly scaled).

The Southern California Earthquake Center Broadband Platform (SCEC BBP; \cite{Maechling2015}) is used for simulating the accelerograms. Of the available simulation methods, the UCSB method \citep{Crempien2015}
is used as it allows $f_c$ to be specified explicitly. It is important to set $f_c$ to the same value for both unscaled and scaled-simulated motions. Only then, as per the Brune's source model, stress drop will scale linearly with seismic moment scaling; see equation \eqref{eqn:Brune} or \eqref{eqn:sourceFin}. Other available simulation methods in the SCEC BBP such as \cite{Graves2010a} and EXSIM \citep{Atkinson2015} pre-specify a constant value for stress drop which might depend upon the tectonic region. As per equation \eqref{eqn:Brune} or \eqref{eqn:sourceFin} then, $f_c$ decreases as the seismic moment scales up linearly. While a reducing $f_c$ with increasing seismic moment is consistent with the self-similarity assumption or the constant stress drop assumption \citep{Boore2003}, it will violate the seismological conditions necessary for simulating a scaled accelerogram that is $\lambda$ times an unscaled one\footnote{It will be helpful here to revisit Figure \ref{fig:freq} in order to recall the influence of $f_c$ on the Fourier spectrum shape.}. Studies have observed variability in earthquake stress drop values (for e.g., see \cite{Allmann2009} or \cite{Baltay2011}), and later on in this paper, this stress drop variability will be utilized to discuss the seismological correctness of accelerogram scaling. 

\subsection{Comparing explicitly scaled and scaled simulated accelerograms}

\begin{figure}[H]
\begin{subfigure}{0.5\textwidth}
\centering
\includegraphics[width=0.85\textwidth]{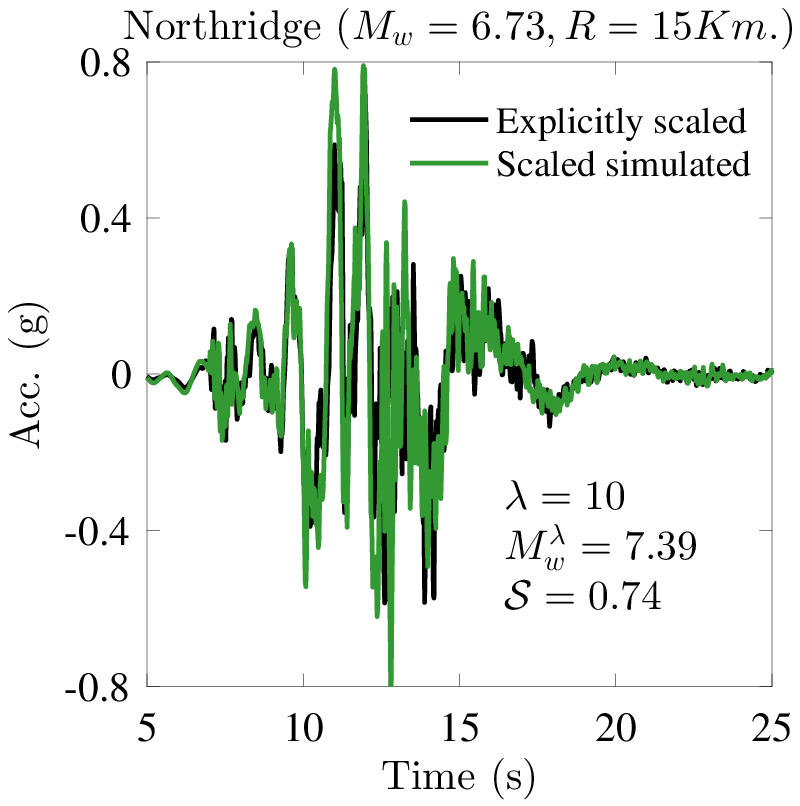}
\caption{} \label{fig:valid1}
\end{subfigure}\hspace*{\fill}
\begin{subfigure}{0.5\textwidth}
\centering
\includegraphics[width=0.85\textwidth]{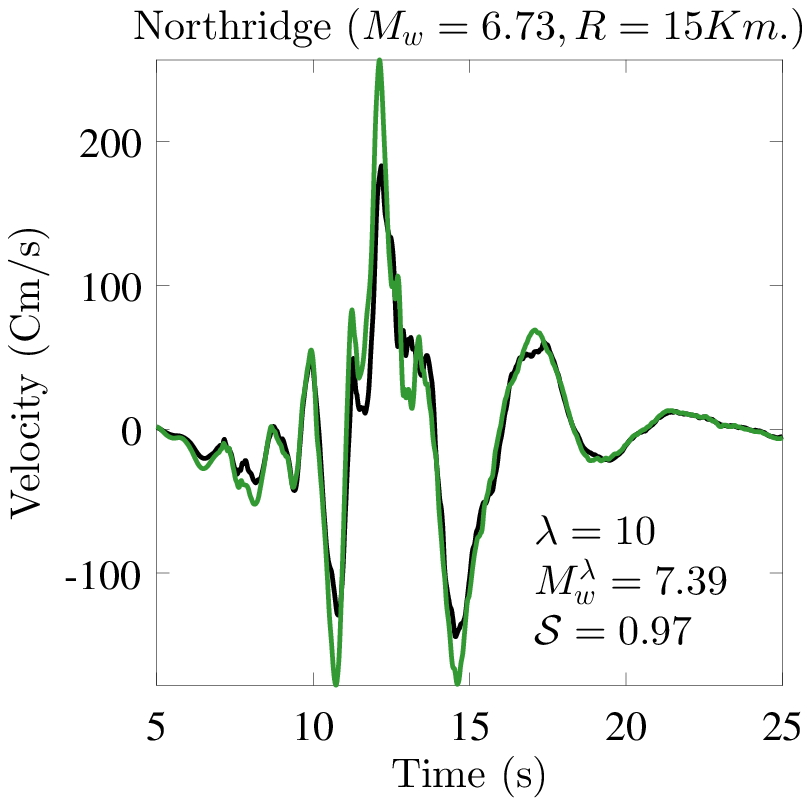}
\caption{} \label{fig:valid2}
\end{subfigure}
\begin{subfigure}{0.5\textwidth}
\centering
\includegraphics[width=0.85\textwidth]{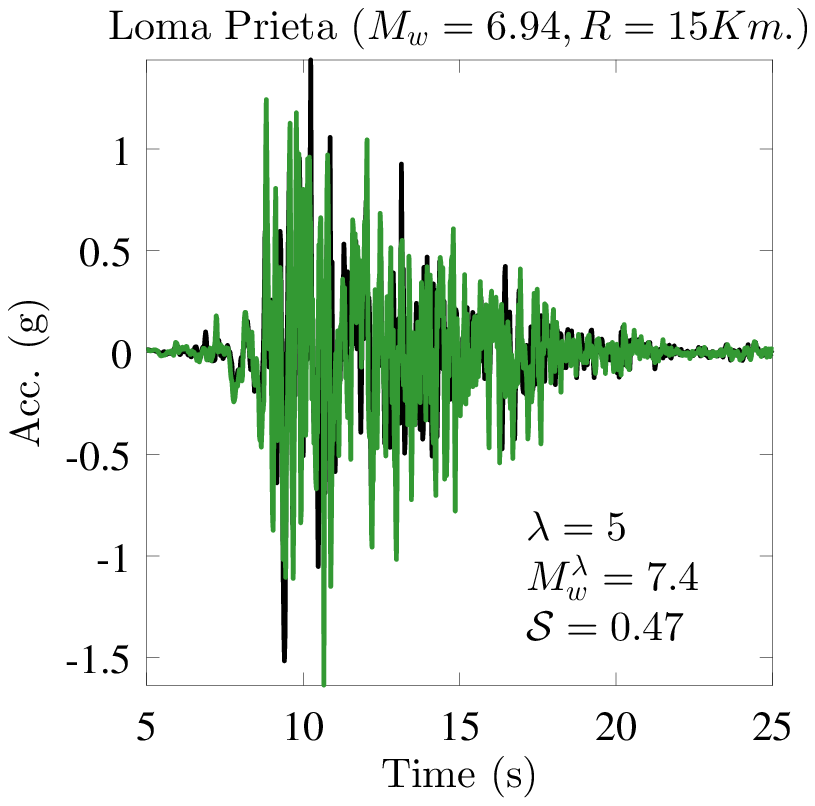}
\caption{} \label{fig:valid3}
\end{subfigure}\hspace*{\fill}
\begin{subfigure}{0.5\textwidth}
\centering
\includegraphics[width=0.85\textwidth]{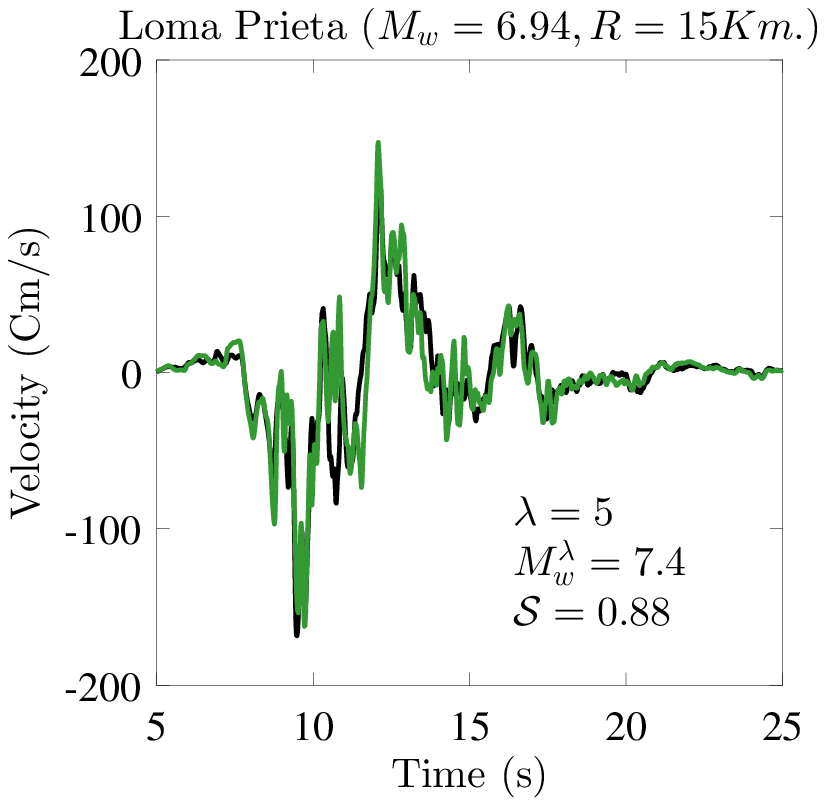}
\caption{} \label{fig:valid4}
\end{subfigure}
\begin{subfigure}{0.5\textwidth}
\centering
\includegraphics[width=0.85\textwidth]{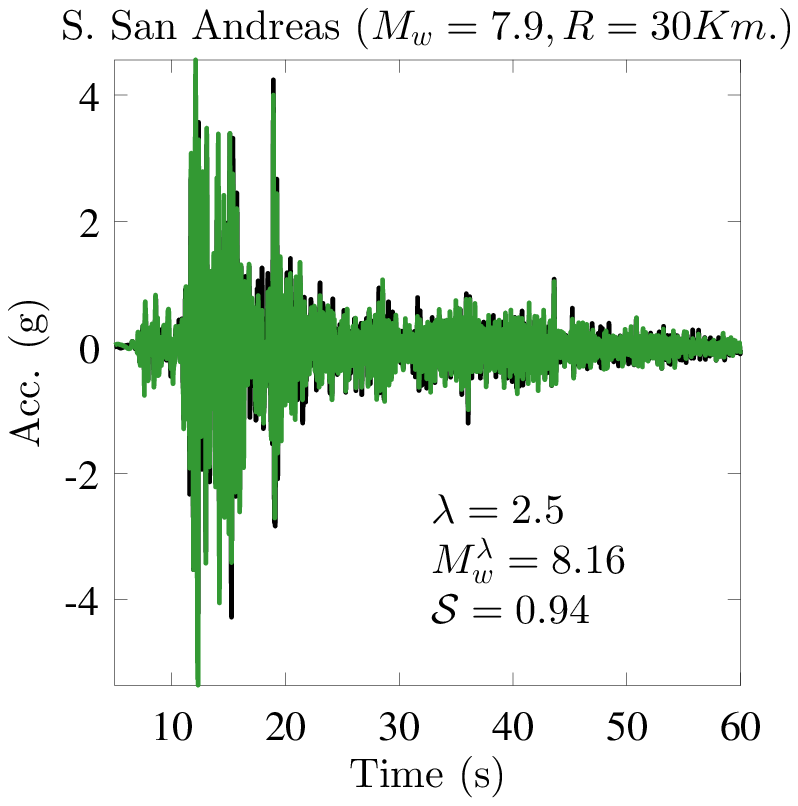}
\caption{} \label{fig:valid5}
\end{subfigure}\hspace*{\fill}
\begin{subfigure}{0.5\textwidth}
\centering
\includegraphics[width=0.85\textwidth]{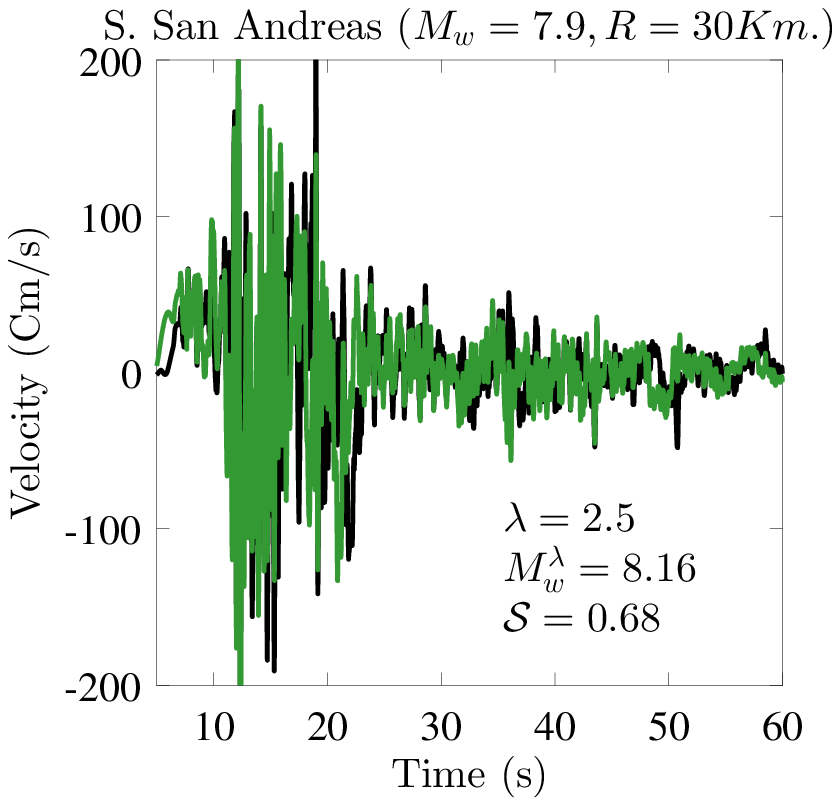}
\caption{} \label{fig:valid6}
\end{subfigure}
\caption{Time domain validation of the seismological interpretation of ground motion scaling considering three earthquake scenarios: Northridge, Loma Prieta, and South San Andreas. Validation is made using simulated ground motions. The black plots are ground motions scaled by explicitly multiplying a scale factor. The green plots are scaled motions implicitly generated using the seismological interpretation developed in this study. The magnitude of the unscaled motions are presented above each sub-figure. The scale factor ($\lambda$) and the magnitude of the scaled motion are presented within each sub-figure. The strict similarity measure $(\mathcal{S})$ values are also shown within each sub-figure.}\label{fig:valid}
\end{figure}

Figure \ref{fig:valid} presents a comparison of the explicitly scaled and the scaled simulated ground motions in the time domain considering the three earthquake scenarios: Northridge, Loma Prieta, and South San Andreas. The magnitude, distance, and scale factor combinations used for these three scenarios are (6.73, 15Km, 10), (6.94, 15Km, 5), and (7.9, 30Km, 2.5). Acceleration and velocity recordings are presented on the left and right panes, respectively, in Figure \ref{fig:valid}. While it is noted that there are some differences in the comparisons provided due to random noise, the explicitly scaled and the scaled simulated motions are seen to be in good general agreement and look very identical. In order to facilitate a stricter comparison between the explicitly scaled and the scaled simulated motions, a similarity metric based on the Normalized Cross-Correlation between two time series is used \citep{Papoulis1962}: 

\begin{equation}
    \label{eqn:sim}
    \mathcal{S} = \max_{\mathcal{L}}~\frac{\int u^\lambda(t)~\widetilde{u}^\lambda(t-\mathcal{L})~dt}{\sqrt{\big[\int u^\lambda(t)~{u}^\lambda(t)~dt\big]\big[\int \widetilde{u}^\lambda(t)~\widetilde{u}^\lambda(t)~dt\big]}}
\end{equation}

\noindent where $u^\lambda(t)$ and $\widetilde{u}^\lambda(t)$ are treated as explicitly scaled and the scaled simulated motions, respectively, and $\mathcal{L}$ is the lag-length. The notation $\mathcal{S}$ is used for indicating that the above metric is a ``strict'' similarity measure and it compares two time series at every discrete time step. It is noted that the value of $\mathcal{S}$ lies between $-1$ and $+1$, and a value equal to $+1$ implies that two time series are the same. Values of $\mathcal{S}$ presented in Figure \ref{fig:valid} are generally seen to be large with the exception of Figure \ref{fig:valid3} where the differences in random noise are noted to be large. Overall, the average value of $\mathcal{S}$ across the six sets of results is $0.78$, and excluding Figure \ref{fig:valid3}, the average is $0.84$.

\begin{figure}[H]
\begin{subfigure}{0.5\textwidth}
\centering
\includegraphics[width=0.85\textwidth]{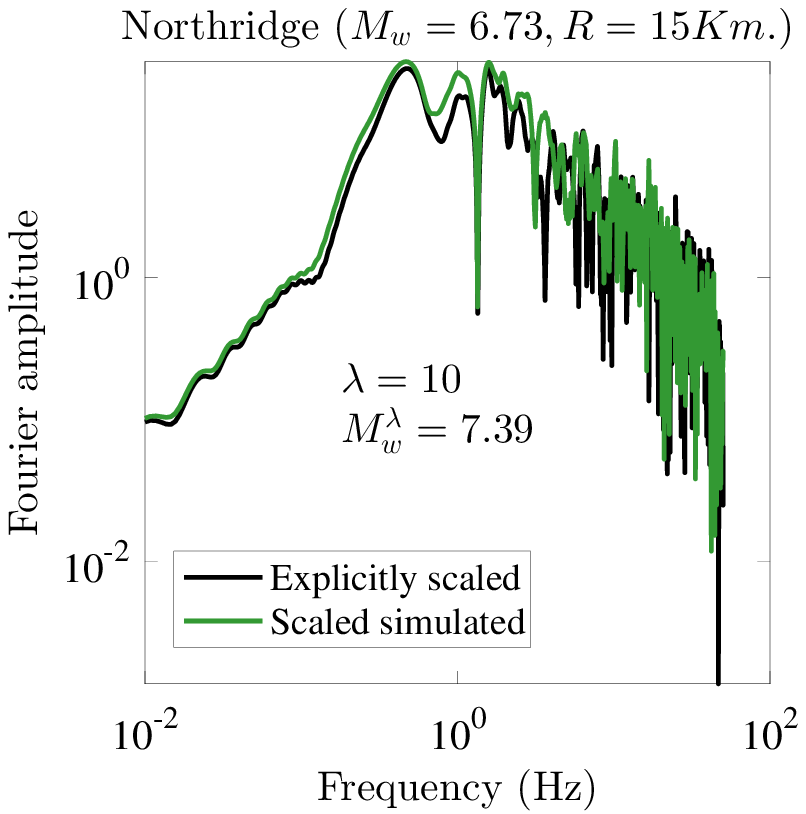}
\caption{} \label{fig:Fvalid1}
\end{subfigure}\hspace*{\fill}
\begin{subfigure}{0.5\textwidth}
\centering
\includegraphics[width=0.85\textwidth]{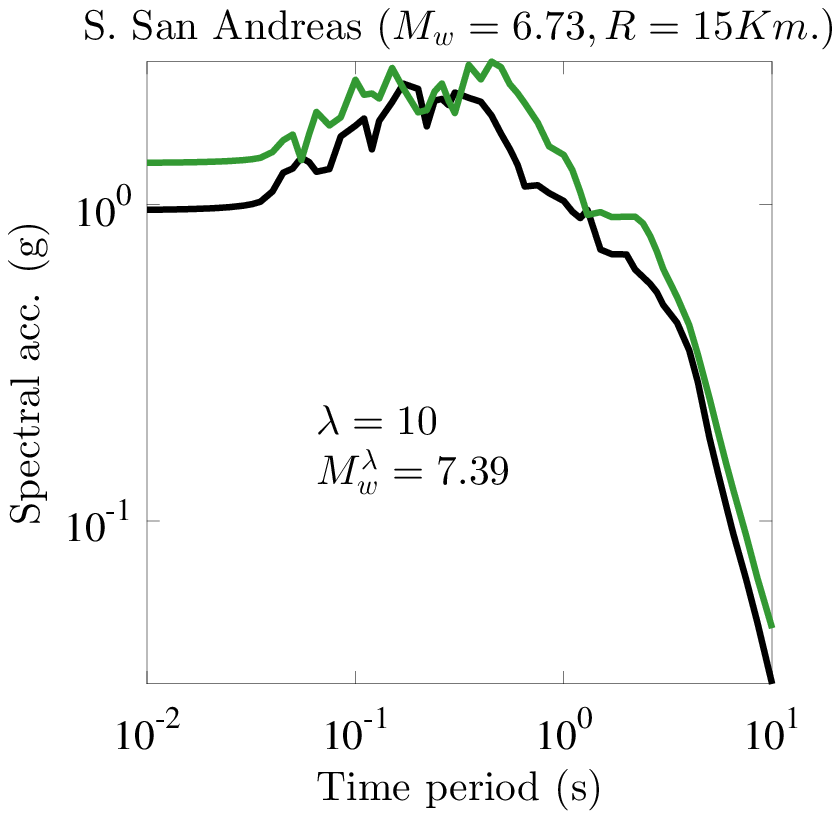}
\caption{} \label{fig:Fvalid2}
\end{subfigure}
\begin{subfigure}{0.5\textwidth}
\centering
\includegraphics[width=0.85\textwidth]{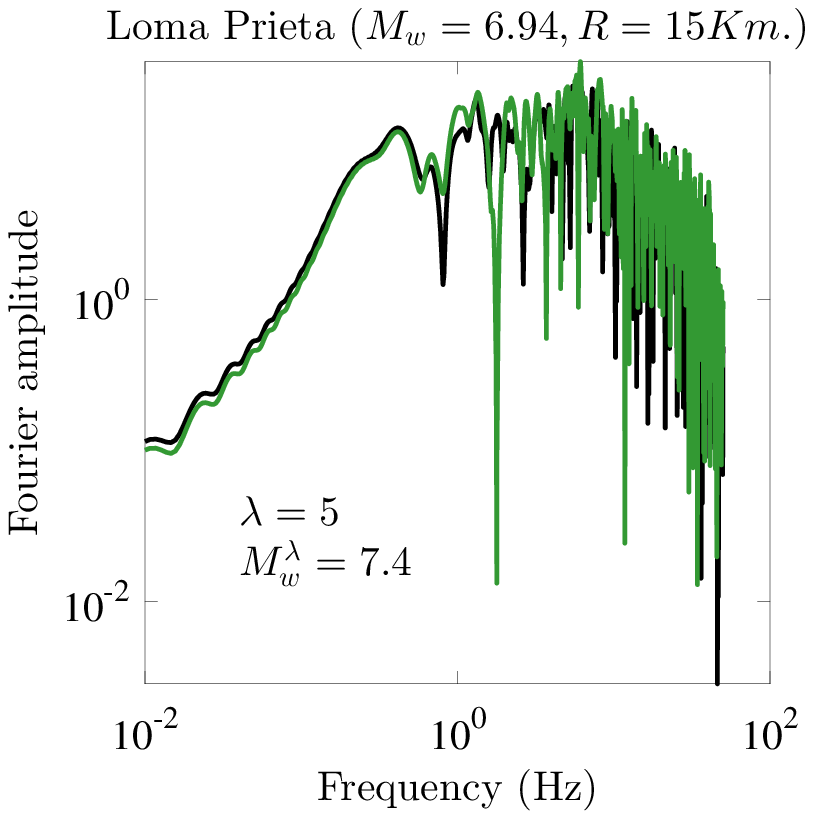}
\caption{} \label{fig:Fvalid3}
\end{subfigure}\hspace*{\fill}
\begin{subfigure}{0.5\textwidth}
\centering
\includegraphics[width=0.85\textwidth]{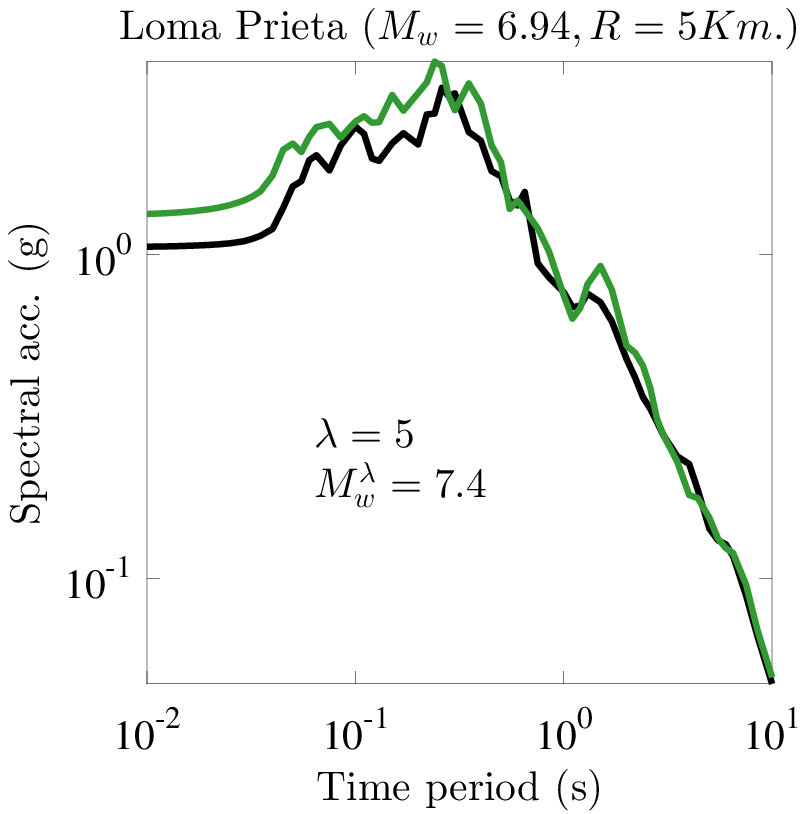}
\caption{} \label{fig:Fvalid4}
\end{subfigure}
\begin{subfigure}{0.5\textwidth}
\centering
\includegraphics[width=0.85\textwidth]{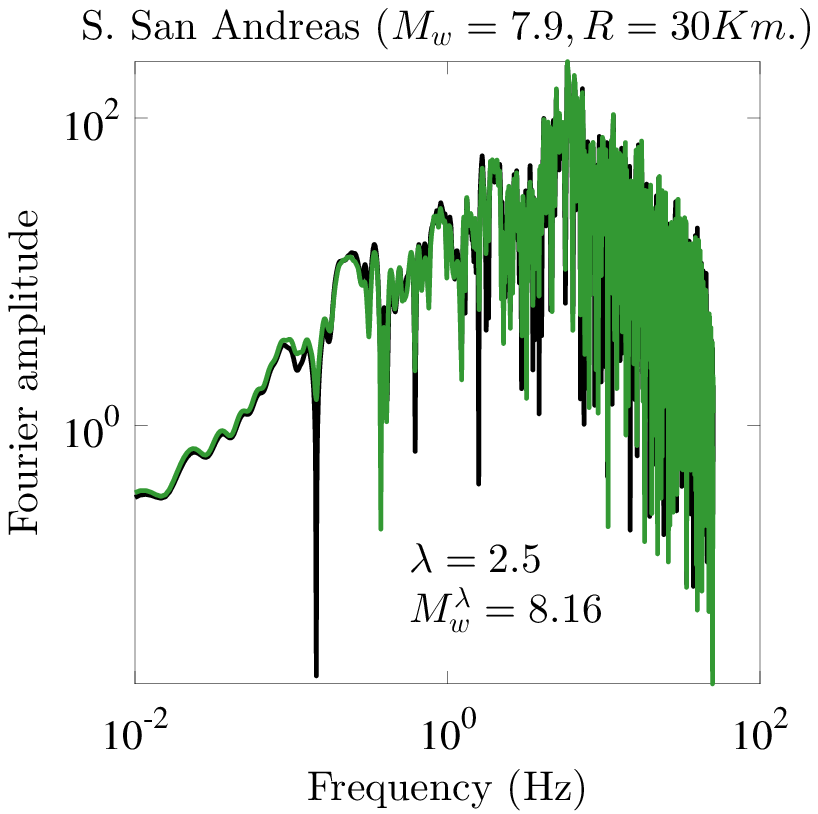}
\caption{} \label{fig:Fvalid5}
\end{subfigure}\hspace*{\fill}
\begin{subfigure}{0.5\textwidth}
\centering
\includegraphics[width=0.85\textwidth]{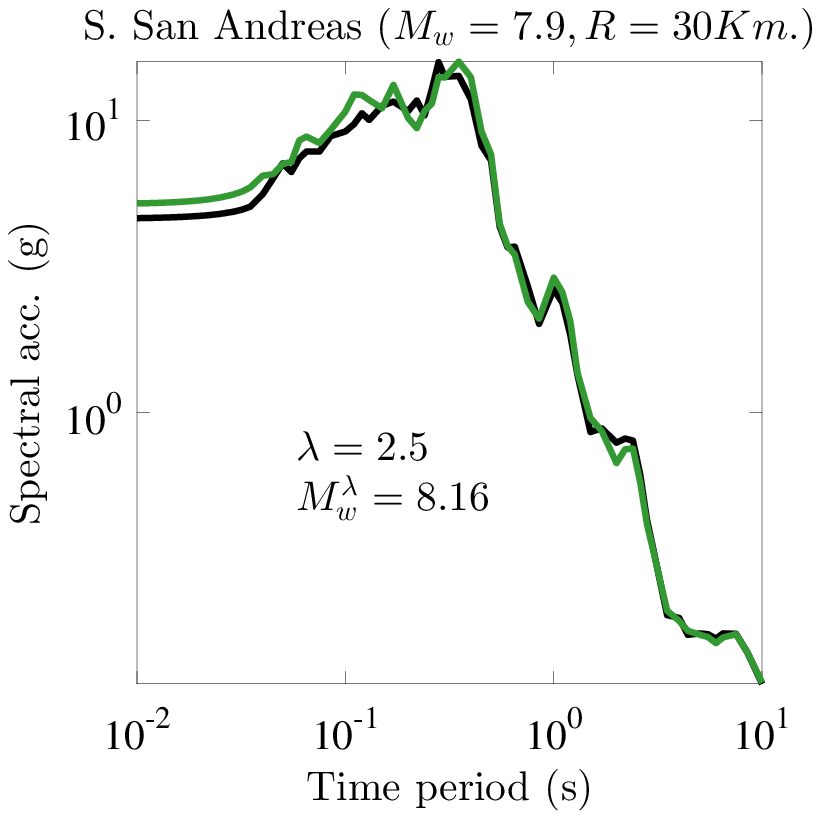}
\caption{} \label{fig:Fvalid6}
\end{subfigure}
\caption{Frequency domain validation of the seismological interpretation of ground motion scaling considering three earthquake scenarios: Northridge, Loma Prieta, and South San Andreas. Validation is made using simulated ground motions. The black plots are ground motions scaled by explicitly multiplying a scale factor. The green plots are scaled motions implicitly generated using the seismological interpretation developed in this study. The magnitude of the unscaled motions are presented above each sub-figure. The scale factor ($\lambda$) and the magnitude of the scaled motion are presented within each sub-figure.}\label{fig:Fvalid}
\end{figure}

Figure \ref{fig:Fvalid} presents a comparison of the explicitly scaled and the scaled simulated ground motions in the frequency domain for the same three earthquake scenarios and the same magnitude, distance, and scale factor combinations as before. Fourier and Response spectra are presented on the left and right panes, respectively, in Figure \ref{fig:Fvalid}. The results are again observed to be in good general agreement, which may be expected because the time domain validations were quite successful.   

\section{Discussion}

Two aspects of the proposed seismological interpretation of accelerogram scaling are additionally discussed. The first one is its applicability to multiple sites under the same set of scaled seismic parameters. And the second one is a discussion over whether linear scaling of stress drop and logarithmic scaling of magnitude under a fixed rupture area is seismologically correct or not. 

\subsection{Applicability of the proposed interpretation of scaling to multiple sites}

The proposed interpretation of accelerogram scaling modifies two seismic source parameters $M_w$ and $\Delta \sigma$ by scale factor $\lambda$. All other parameters such as rupture area and corner frequency, and conditions such as seismic wave attenuation and site response effects are set to be the same for both the unscaled and the scaled accelerograms. Due to modifying source parameters only, the proposed interpretation of accelerogram scaling is applicable to multiple sites. That is, if an accelerogram at site $\mathbf{X}$ is scaled by $\lambda$, the proposed interpretation suggests that an accelerogram recorded at any other site $\mathbf{X}_1~(\mathbf{X}_1 \neq \mathbf{X})$ is also scaled by $\lambda$. This physically seems plausible because when an accelerogram at $\mathbf{X}$ is scaled, some aspects of the seismic source are modified and these modifications also influence the accelerograms recorded at some other site $\mathbf{X}_1$. 

\begin{figure}[h]
\begin{subfigure}{0.5\textwidth}
\centering
\includegraphics[width=0.85\textwidth]{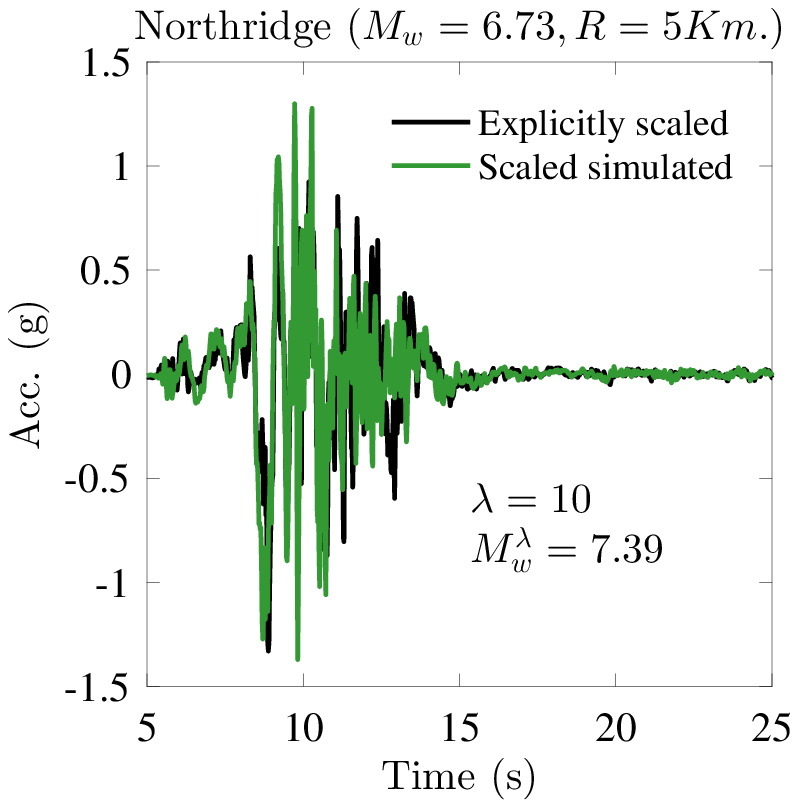}
\caption{} \label{fig:Sites1}
\end{subfigure}\hspace*{\fill}
\begin{subfigure}{0.5\textwidth}
\centering
\includegraphics[width=0.85\textwidth]{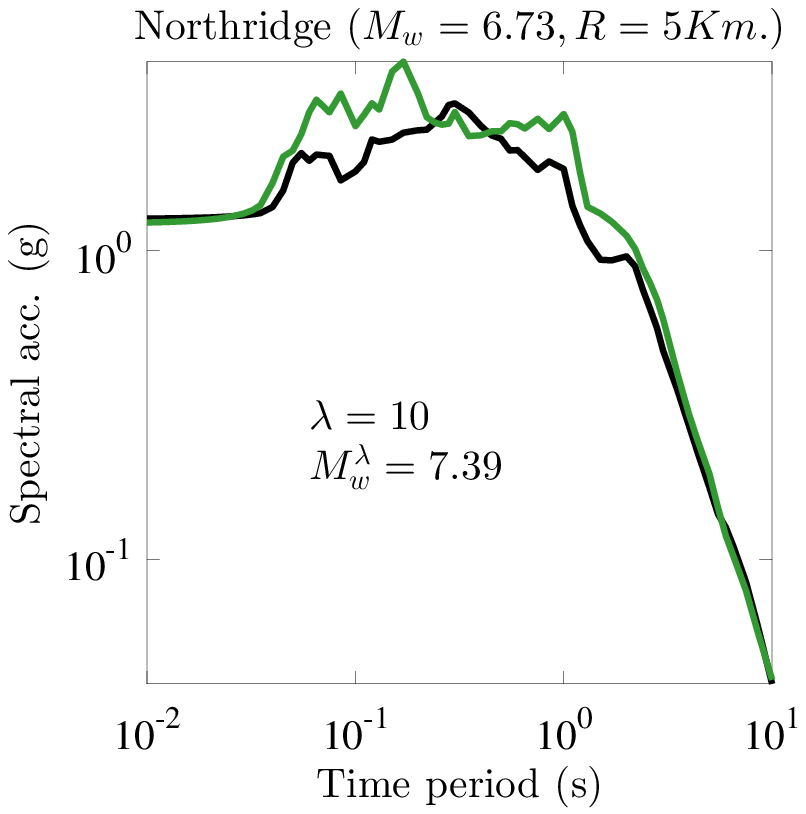}
\caption{} \label{fig:Sites2}
\end{subfigure}
\caption{Applicability of the proposed interpretation of accelerogram scaling to multiple sites. Explicitly scaled and scaled simulated (a) Accelerograms and (b) spectral accelerations from the Northridge earthquake recorded at a distance of 5Km (as opposed to 15Km used in Figures \ref{fig:valid1} and \ref{fig:Fvalid2}).}\label{fig:Sites}
\end{figure}

Figure \ref{fig:Sites} presents an example demonstrating the applicability of the proposed interpretation of scaling to multiple sites. In this example, all the conditions to simulate the unscaled and the scaled Northridge accelerograms are fixed to be the same as previous (see Figure \ref{fig:valid1}), except that these accelerograms are recorded at $5$Km distance instead of $15$Km. Figures \ref{fig:Sites1} and \ref{fig:Sites2}, presenting the accelerograms and spectral accelerations, respectively, suggest that the proposed interpretation of scaling is valid even at this new site.


\subsection{Seismological correctness of accelerogram amplitude scaling}

A question that has interested both seismologists and earthquake engineers alike concerns the seismological correctness of accelerogram scaling. In this paper, it was proposed that linearly scaling an accelerogram scales $\Delta \sigma$ linearly and $M_w$ logarithmically. Among the other seismic parameters, it was noted that the effective rupture area $\mathcal{A}$ is invariant to scaling an accelerogram. It is further suggested that the physical admissibility of modifying $\Delta \sigma$ and $M_w$ in this manner may not be treated as a binary question. The reason is, given $\mathcal{A}$, $M_w$ and $\Delta \sigma$ are empirically observed to be uncertain. As presented below, probability theory can be used to quantify this uncertainty:

\begin{equation}
    \label{eqn:M_SD}
    \begin{aligned}
    p(M_{res},\Delta \sigma|\mathcal{A}) &\approx p(M_{res}|\mathcal{A})~p(\Delta \sigma|\mathcal{A})\\
    & \approx p(M_{res}|\mathcal{A})~p(\Delta \sigma)\\
    \end{aligned}
\end{equation}

\noindent where $p(.|.)$ denotes a conditional probability distribution and $M_{res}$ is the magnitude residual obtained by taking a difference between the predicted and the observed magnitudes given $\mathcal{A}$. It is noted in the above equation that the independence between $M_w$ and $\Delta \sigma$ \citep{Allmann2009,Baltay2011} is used to split the joint conditional probability $p(M_{res},\Delta \sigma|\mathcal{A})$ into marginals $p(M_{res}|\mathcal{A})$ and $p(\Delta \sigma|\mathcal{A})$. Furthermore, in line with the general notion in seismology, $\Delta \sigma$ is assumed to be independent of $\mathcal{A}$. 

\begin{figure}[h]
\begin{subfigure}{0.5\textwidth}
\centering
\includegraphics[width=0.85\textwidth]{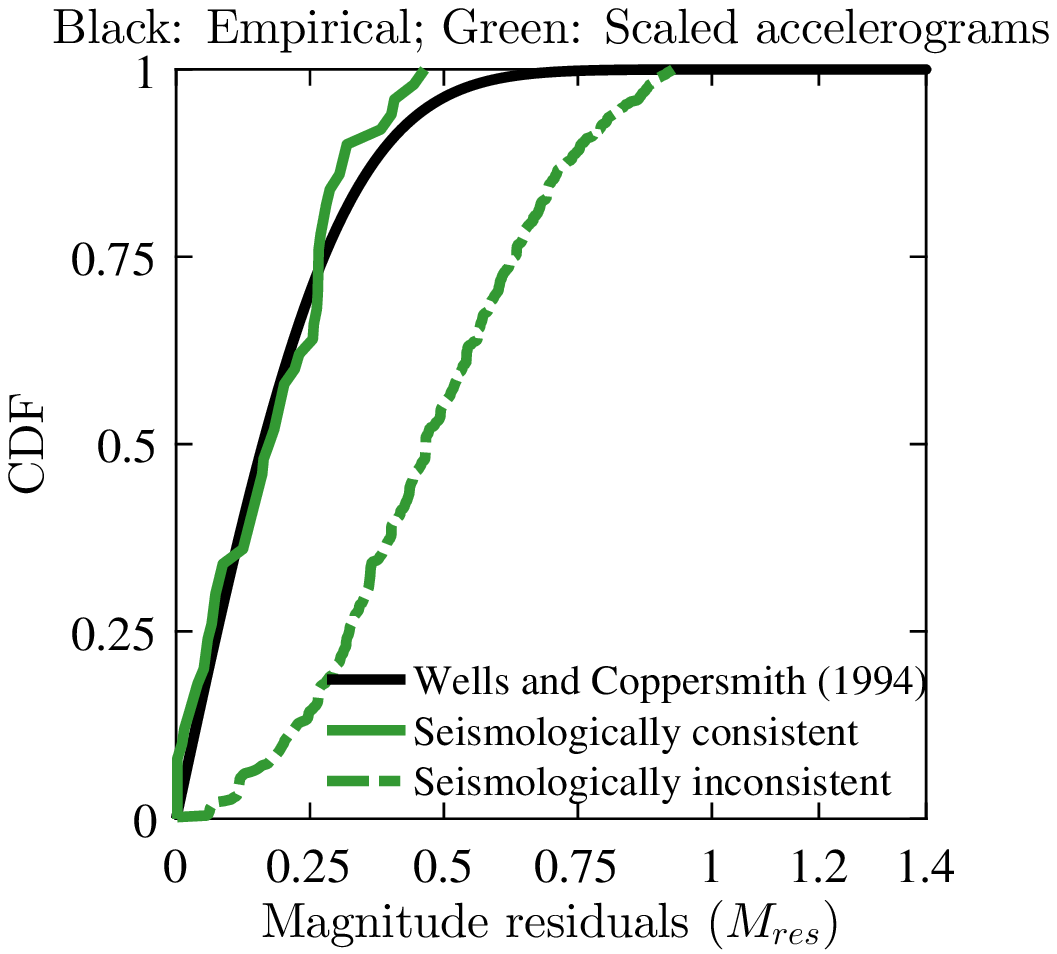}
\caption{} \label{fig:ConsSeis1}
\end{subfigure}\hspace*{\fill}
\begin{subfigure}{0.5\textwidth}
\centering
\includegraphics[width=0.85\textwidth]{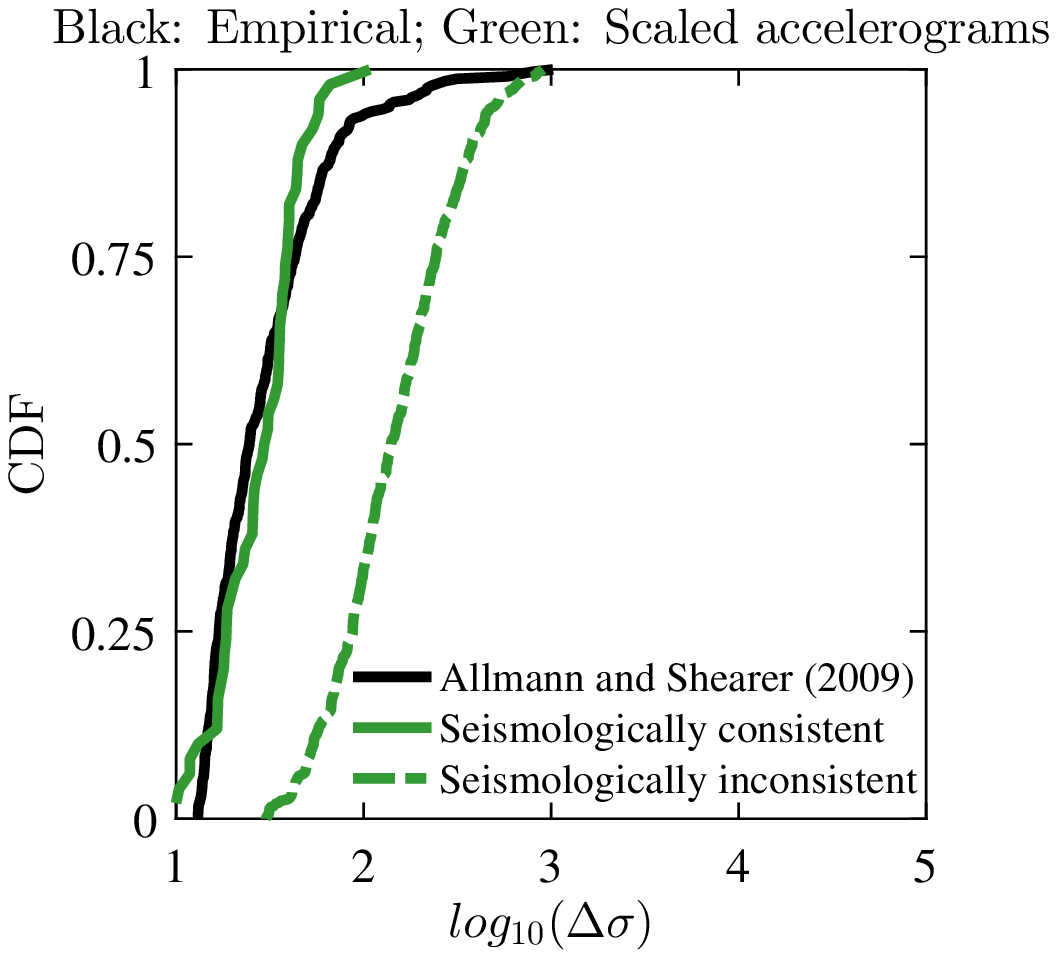}
\caption{} \label{fig:ConsSeis2}
\end{subfigure}
\caption{Depiction of seismological correctness of accelerogram scaling through matching (a) $p(M_{res}|\mathcal{A})$ and (b) $\Delta \sigma$ distributions of the scaled accelerograms with empirical observations. It is noted that while a suite of scaled accelerograms (represented by solid green plots) obey the empirically observed $M_{res}|\mathcal{A}$ and $\Delta \sigma$ distributions, another scaled suite (represented by dashed green plots) does not. Accelerograms here are assumed to be scaled upwards (i.e., $\lambda >1$).}\label{fig:ConsSeis}
\end{figure}

Through the independence of $M_{res}|\mathcal{A}$ and $\Delta \sigma$ distributions demonstrated by equation \eqref{eqn:M_SD}, it is suggested that one way to scale accelerograms in a seismologically correct manner is to ensure that a suite of scaled accelerograms have the same $p(M_{res}|\mathcal{A})$ and $p(\Delta \sigma)$ distributions as empirical observations. These distributions are straightforward to compute for the suite of scaled accelerograms because $\Delta \sigma^\lambda = \lambda~\Delta \sigma$ and $M_{res} = M_w^\lambda - M_w$; rupture area for a scaled accelerogram does not change from what it was for the unscaled one. The empirical $p(M_{res}|\mathcal{A})$ distribution can be computed using the \cite{Wells1994} magnitude-rupture area relationship:

\begin{equation}
    \label{eqn:Wells}
    p(M_{res}|\mathcal{A}) = erf\Big(\frac{M_{res}}{\sigma~\sqrt{2}}\Big)
\end{equation}

\noindent where $M_{res}$ is the difference between predicted and observed magnitudes, $\sigma$ is the standard deviation of the empirical relation, and $erf$ is an error function. It is noticed that the above equation represents the cumulative distribution function of a Half-Normal distribution and is valid when accelerograms are scaled upwards (i.e., $\lambda>1$ and $M_{res} > 0$). The empirical $p(\Delta \sigma)$ distribution can be obtained from \cite{Allmann2009}. Figure \ref{fig:ConsSeis} presents an illustration of seismological correctness of accelerogram scaling from which it is noted that while a suite of scaled accelerograms (represented by solid green plots) obey the empirically observed $M_{res}|\mathcal{A}$ and $\Delta \sigma$ distributions, another scaled suite (represented by dashed green plots) does not.

\section{Summary and Conclusions}

Due to the paucity of recorded accelerograms that are intense enough to cause damage/collapse of structural models, accelerogram amplitude scaling is employed in earthquake engineering analysis. If scaled accelerograms are being used for target spectrum matching, seismic response analysis, and seismic risk assessment, then these accelerograms should represent potential earthquake events that are yet to be realized. What are the magnitudes of such earthquake events that would result in scaled accelerograms? At what distances such earthquake events happen from the recording sites? And what about parameters such as rupture area and stress drop? These are the type of questions this paper has attempted to answer by conducting a theoretical investigation into the seismology of accelerogram scaling.

Representation theorem and accelerogram Fourier spectrum were used to investigate the seismology of scaled accelerograms. The following deductions were made:\begin{itemize}
\item Unscaled and scaled accelerograms have the same spatial as well as temporal distribution of rupture over the fault plane. The rupture amplitudes (given a position or a time instant at the fault) of the scaled accelerogram, however, are scaled by factor $\lambda$. Consequently, magnitude of a scaled accelerogram becomes: $M_w^\lambda = M_w+2/3~log_{10}(\lambda)$
\item The effective rupture dimensions are the same for unscaled and scaled accelerograms. These accelerograms are further recorded at the same site, leading the Joyner-Boore distance metric to be invariant to accelerogram scaling.
\item Unscaled and scaled accelerograms have similar Fourier spectrum shape. This necessitates the corner frequency of these accelerograms to be the same, leading the static stress drop to scale linearly with accelerogram scaling ($\Delta \sigma^\lambda = \lambda~\Delta \sigma$).
\end{itemize}
\noindent Additionally, it should be noted that the soil underlying a site should transmit the bedrock motion to the surface in the same manner for both unscaled and scaled accelerograms; that is, relative nonlinearity of site response between these accelerograms must not be large.  

The proposed seismological interpretation of accelerogram amplitude scaling was validated using the UCSB hybrid method for ground motion simulation by comparing the explicitly scaled and the scaled (implicitly) simulated motions. Three earthquake scenarios were considered: Northridge, Loma Prieta, and South San Andreas. In the time domain, these ground motions are compared using a similarity measure $\mathcal{S} \in [-1,~+1]$ (higher value desirable). Across six sets of accelerogram and velocity recordings, the explicitly scaled and the scaled simulated motions were found to be in good agreement, and the average value of $\mathcal{S}$ was found to be 0.78. These motions were also found to compare well in the frequency domain, particularly under representations such as the Fourier amplitude spectrum and the response spectrum.

A key feature of the proposed interpretation of scaling is its recommendation that if an accelerogram at a site is scaled by $\lambda$, accelerograms recorded at other sites are also scaled by $\lambda$. This is because, accelerogram scaling modifies the source parameters thereby influencing the realized the ground motions at all the sites surrounding a seismic source. Finally, a discussion was made on the seismological correctness of scaling. Ensuring that a suite of scaled accelerograms has magnitude given rupture area and stress drop distributions similar to empirical observations is suggested as a way to scale accelerograms in a seismologically consistent manner.

\bibliography{main.bib}

\end{document}